\documentclass{article}
\usepackage{iclr2027_conference,times}

\usepackage{amsmath,amsfonts,bm}

\def\eqref#1{equation~\ref{#1}}

\def\1{\bm{1}}

\DeclareMathAlphabet{\mathsfit}{\encodingdefault}{\sfdefault}{m}{sl}
\SetMathAlphabet{\mathsfit}{bold}{\encodingdefault}{\sfdefault}{bx}{n}

\usepackage{hyperref}
\usepackage{url}

\usepackage[table]{xcolor}
\usepackage{amsmath}
\usepackage{booktabs}
\usepackage{arydshln}
\usepackage{graphicx}
\usepackage{multicol,multirow}
\usepackage{wrapfig}
\usepackage{subcaption}
\usepackage{pifont}
\usepackage{amssymb}
\usepackage{fdsymbol}

\newcommand{\cmark}{\ding{51}}
\newcommand{\xmark}{\ding{55}}

\newcommand{\method}{\textsc{SEA-LM}}
\newcommand{\samethod}{\textsc{FOACoder}}
\newcommand{\samethodfull}{\textsc{FOACoder: \underline{F}irst \underline{O}rder \underline{A}mbisonics En\underline{coder}}}
\newcommand{\methodfull}{\textsc{SEA-LM: \underline{S}patio-Temporal \underline{E}gocentric \underline{A}udio \underline{L}anguage \underline{M}odel}}
\newcommand{\basemodel}{\textsc{Gemma-4-E4B-it}}
\definecolor{methodrow}{RGB}{225,240,250}

\title{SEA-LM: Egocentric Spatial Audio Understanding for Wearable Microphone Arrays}

\author{
\parbox{0.98\textwidth}{\centering\normalfont
\textbf{Sonal Kumar}$^{1,\dagger,\ast}$
\quad \textbf{Sinan Hersek}$^{2,\ddagger,\ast}$
\quad \textbf{Artem Dementyev}$^{3}$
\quad \textbf{Mengzhen Pan}$^{2}$\\[0.4ex]
\textbf{Ishan Chatterjee}$^{2}$
\quad \textbf{Anurag Kumar}$^{3}$
\quad \textbf{Ramani Duraiswami}$^{1}$
\quad \textbf{Dinesh Manocha}$^{1}$
\quad \textbf{Andrea Colaco}$^{2,\S}$\\[0.6ex]
$^{1}$University of Maryland, College Park
\quad $^{2}$Google
\quad $^{3}$Google DeepMind\\[0.4ex]
\textit{Corresponding authors:} \texttt{sonalkum@umd.edu, shersek@google.com}
}}

\iclrfinalcopy
\begin{document}

\maketitle

\begingroup
\renewcommand{\thefootnote}{\fnsymbol{footnote}}
\footnotetext[1]{Equal contribution.}
\footnotetext[2]{Work done while at Google XR.}
\footnotetext[3]{Now at Google DeepMind, work done while at Google XR.}
\footnotetext[4]{Now at Meta, work done while at Google XR.}
\endgroup

\begin{abstract}
Embodied, ego-centric intelligence fundamentally requires the ability to comprehend spatial audio within complex environments. While large audio-language models excel at mono-channel reasoning, they lack spatial awareness, discarding critical spatial cues that enable sound localization and that can improve the disentanglement of overlapping sound sources. To address this, we present \method{}, a Spatial Audio Understanding model. First, we introduce \samethod{}, a layout-flexible spatial audio encoder trained on source localization and ego-centric voice activity detection objectives to encode First Order Ambisonics derived from variable-count, variable-position smart-glasses arrays via beamforming. We train a Multimodal Large Language Model (MLLM) to understand these spatial audio embeddings through a two-stage curriculum spanning six tasks, including sound localization and spatially selective transcription in settings with multiple speakers and overlapping sounds. To prevent the transcription outputs from dominating the next token prediction loss and overwhelming the direction predictions, we introduce a \textit{Spatio-temporal Weighted Cross-Entropy Loss}. On our evaluation set, \method{} achieves lower azimuth and elevation MAE, higher temporal IoU, lower external-source hallucination and missing-source rates, and lower WER on most transcription tasks than compared baselines, while remaining robust across 1,211 smart-glasses array configurations with 4 to 9 microphones.
\end{abstract}

\section{Introduction}
\label{sec:intro}
The advent of Large Audio-Language Models \citep{chu2023qwen, tang2023salmonn, gong2023ltu-as, af3} built on robust audio encoders \citep{radford2023robust,elizalde2022clap,wu2023large,chen2022beats,ast} has brought a paradigm shift to machine audition, enabling complex semantic reasoning and open-ended conversational tasks. However, these systems are built on understanding monophonic audio discarding spatial information in the soundscape. In both human psychoacoustics and machine audition, the spatial cues contained in spatial audio (e.g., inter-channel time and level differences) have been shown to provide significant advantages in speech intelligibility \citep{litovsky2012spatial, hawley2004benefit} and the disentanglement of overlapping speakers in complex ``cocktail party'' environments \citep{wang2018combining, chen2018multi}. Spatial audio understanding is increasingly critical for emerging embodied platforms such as smart glasses, social robots, and autonomous vehicles, where multi-channel microphone arrays are increasingly more common and accurate 3D sound source localization enables robust scene awareness, natural human-machine interaction, and safe navigation in dynamic environments.

While recent generalized multi-channel audio foundation models (e.g., ELSA \citep{devnani2024learning} and M-BEST-RQ \citep{yang2025m}) learn versatile spatial representations, they have yet to be fully integrated with modern MLLMs to demonstrate broad spatial audio understanding. To bridge this gap, recent works have injected spatial awareness directly into MLLMs, though they face limitations regarding hardware generalization and data realism. Binaural models like OWL \citep{owl2025} and Dspast \citep{dspast2025} are restricted to earbud-like form factors, and extending them to arbitrary microphone arrays requires pre-processing that degrades spatial cues \citep{madmoni2025design}. First Order Ambisonics (FOA) models like Sci-Phi \citep{sciphi2025} and The World is Not Mono \citep{worldnotmono2025} are trained on idealized synthetic Ambisonic room impulse responses or data from expensive, idealized spherical arrays (e.g., STARSS23 \citep{shimada2023starss23}) rather than ubiquitous, scalable devices like smart glasses. By omitting the artifacts introduced when converting audio from arbitrary device geometries into FOA, these models remain specialized to idealistic FOA spatial audio.

Furthermore, these spatial MLLMs generally lack a true ego-centric perspective, failing to jointly model wearer speech (e.g., the speech of a glasses wearer or the self-speech of a robot head) alongside external sounds. They are also prone to \textit{token dilution}, where dense speaker transcripts dominate the next-token prediction loss, overwhelming the sparse direction and timestamp predictions.

\noindent {\bf Main Results:} To overcome these limitations, we propose \method{}, a comprehensive Spatial Audio Understanding framework. First, we introduce \samethod{}, a layout-flexible encoder pre-trained on wearer Voice Activity Detection (VAD) and Sound Event Localization and Detection (SELD) objectives to map First Order Ambisonics (FOA) from variable-count and variable-position smart-glasses arrays into spatial audio embeddings.
Second, we introduce our MLLM, \method{}, that combines these spatial audio embeddings with embeddings from a standard mono audio encoder to enable ego-centric spatial audio understanding. Trained via a two-stage curriculum spanning six tasks, \method{} jointly comprehends \textit{what} sounds are present, \textit{where} they originate, \textit{when} they occur, \textit{what} is being said, and \textit{who} is speaking (i.e., the wearer versus bystanders). This explicit differentiation between wearer and bystander speech enables true ego-centric scene understanding. Furthermore, to combat token dilution, we introduce a \textit{Spatio-temporal Weighted Cross-Entropy Loss} that dynamically amplifies the training signal on both directional and temporal tokens.
On our evaluation set, \method{} reduces azimuth, elevation, and distance MAE, improves temporal IoU and query F1, lowers external-source hallucination and missing-source rates, and reduces WER on most transcription tasks relative to the baselines.

\noindent In summary, our key contributions are:

\noindent\textbf{(i) Spatial Audio Simulation and Encoding Methodology:} We introduce a data generation and encoding pipeline that incorporates the realistic device physics of variable-count and variable-position head-worn microphone arrays alongside the preprocessing for converting these signals to FOA. While \samethod{} is trained and evaluated on 1,211 configurations selected from two smart-glasses geometries with realistic head and device scattering, our generation methodology is adaptable to other hardware, such as smartphones and robots. Furthermore, by explicitly integrating the Ambisonic Signal Matching (ASM) beamforming step, we expose the encoder and downstream model to the inherent non-idealities of transforming multichannel device audio to FOA.

\noindent\textbf{(ii) Ego-Centric Spatial Audio Model (\method{}):} We introduce an MLLM that jointly performs acoustic, semantic, and spatial reasoning while distinguishing wearer speech from bystander speech in complex, multi-source environments.

\noindent\textbf{(iii) Spatio-Temporal Weighted Optimization:} We introduce a weighted cross-entropy objective that prioritizes sparse direction and timestamp tokens during SFT, mitigating token dilution from dense transcripts.

\noindent\textbf{(iv) Large-Scale Smart-Glasses Evaluation:} We evaluate \method{} without array-specific retraining across 1,211 configurations derived from two realistic smart-glasses geometries, spanning 4 to 9 microphones and varied microphone positions. Relative to the SELDNet-based spatial baseline on Task~1, \method{} reduces azimuth MAE from $80.6^\circ$ to $14.9^\circ$ and elevation MAE from $27.4^\circ$ to $12.8^\circ$, while doubling location-bin accuracy from $42.0\%$ to $84.1\%$. Performance also remains stable across microphone-count cohorts, demonstrating the practical benefit of layout-flexible spatial encoding for smart glasses.

\section{Related Work}
\label{sec:related_work}

\noindent\textbf{Large Audio-Language Models.}
The progress of LLMs has catalyzed the development of Audio Language Models (ALMs). While encoder-only ALMs~\citep{elizalde2022clap, wu2021wav2clip} excel at cross-modal retrieval, encoder-decoder architectures (MLLMs)~\citep{gong2023ltu, tang2023salmonn, chu2023qwen, af2} augment LLMs with dedicated audio encoders to achieve exceptional proficiency in ASR, captioning, and open-ended QA. However, their reliance on mono-channel acoustic inputs discards vital spatial cues (e.g., ITD and ILD). This restriction renders current MLLMs incapable of localizing sound sources or estimating distances, and degrades their ability to disentangle overlapping speakers in complex 3D environments.

\noindent\textbf{Spatial Audio MLLMs.}
Existing spatial-audio MLLMs each address only part of the ego-centric scene-understanding problem. BAT~\citep{bat2024} and OWL~\citep{owl2025} operate on binaural, one- or two-source scenes and do not predict event boundaries, transcribe speech, or distinguish wearer speech from bystanders. SPUR~\citep{sakshi2025spur} consumes FOA, but its benchmark primarily evaluates high-level, qualitative questions about source intent, causal associations, navigation, and mental rotation rather than explicit source characterization; it does not jointly evaluate precise 3D localization, temporal event boundaries, or source-resolved transcription. Sci-Phi~\citep{sciphi2025} provides the closest holistic description, but assumes native FOA and stationary sources and does not model wearable-array conversion artifacts, ego-centric speaker roles, or query-conditioned transcription. PhaseCoder~\citep{dementyev2026phasecoder} supports arbitrary raw arrays and targeted transcription, but its LLM is trained on non-overlapping speech-only scenes, treats non-speech sounds as distractors, and does not model device acoustics. In contrast, \method{} jointly performs continuous 3D localization, temporal grounding, speech and non-speech understanding, overlapping-speech disentanglement, wearer--bystander attribution, and holistic and targeted transcription across realistic smart-glasses arrays.

\section{Methodology}
\label{sec:method}
\begin{figure*}[t]
    \centering
    \includegraphics[width=\linewidth, trim= 0 0.75cm 0 0]{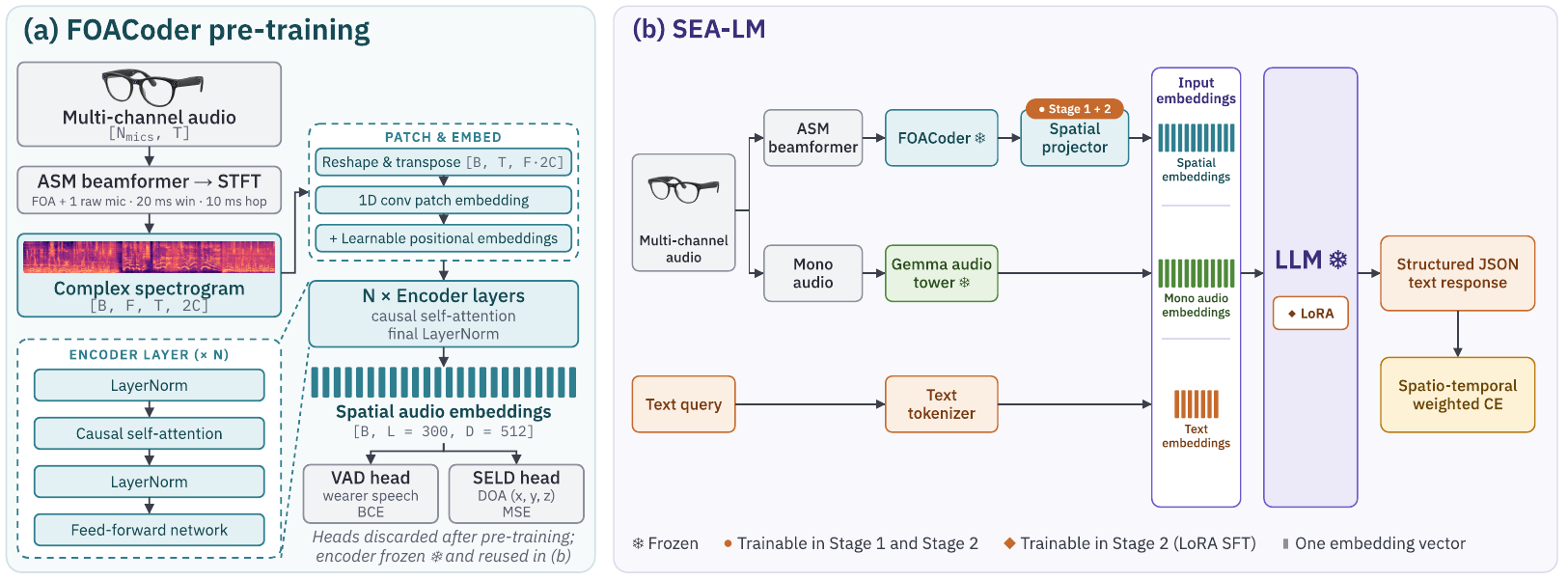}
\caption{\small\textbf{Overview of \method{} and \samethod{} architectures.} (Left) \samethod{}: a causal transformer encoder ingests a multi-channel complex spectrogram and produces encoded spatial tokens. (Right) \method{}: a dual-pathway MLLM that combines frozen \samethod{} spatial features with a frozen semantic audio encoder, injecting both into an LLM for joint spatial and semantic reasoning over FOA audio and text queries. Tensor dimensions use $B$ for batch size, $F$ for STFT frequency bins, $T$ for time frames, $C$ for input audio channels, $L$ for the number of temporal patches or tokens, and $D$ for the embedding dimension; $P$ denotes the temporal patch length and $N$ the number of transformer blocks.}
\label{fig:spatial_MLLM}
\vspace{-1em}
\end{figure*}

Figure~\ref{fig:spatial_MLLM} provides an overview of the complete architecture: \samethod{} converts the multichannel spectrogram into spatial audio embeddings, which \method{} combines with native monaural audio and text embeddings for joint spatial and semantic reasoning. We discuss each component of the model in the following subsections.

\subsection{\samethodfull{}}
\label{subsec:sa_encoder}

To effectively capture spatial cues we introduce \samethod, a dedicated spatial audio encoder operating on First-Order Ambisonics (FOA). Unlike traditional semantic audio encoders that heavily pool temporal dimensions \citep{ast}, our encoder is designed to preserve fine-grained temporal and spatial resolution necessary for precise 3D localization. We use the wearer-centric coordinate convention defined in Appendix~\ref{appendix:cord_sys} throughout.

\subsubsection{Pre-processing}
\label{subsubsec:pre_processing}

To process multichannel audio captured from a selected smart-glasses array, we employ Ambisonic Signal Matching (ASM) beamforming with weights derived for that configuration to transform its variable-count, variable-position channels into a common First-Order Ambisonics format \citep{gayer2024ambisonics}. Given a 30-second time-domain multichannel audio signal $\mathbf{X}_{\text{raw}} \in \mathbb{R}^{N_{\text{mics}} \times T}$ at a 16 kHz sample rate, we transform it into FOA $\mathbf{X}_{\text{foa}} \in \mathbb{R}^{4 \times T}$. Further details regarding this transformation are provided in Appendix \ref{appendix:asm_beamforming_details} that explains the ASM beamforming.

ASM beamforming introduces spectral changes and artifacts such as low-pass filtering, which can result in information loss. To mitigate this information loss, we concatenate one randomly selected raw microphone channel to the FOA audio, forming a 5-channel audio input. Further details on these artifacts can be found in Appendix \ref{appendix:beamforming_artifacts}. We then extract features by computing the Short-Time Fourier Transform (STFT) of this combined signal using a window size of 320 (20 ms), a hop size of 160 (10 ms), and a Hann window. This process results in a complex-valued STFT tensor of shape $F \times T \times C$, where $F=161$ represents the number of frequency bins, $T=3000$ denotes the time frames, and $C=5$ represents the channels.

\subsubsection{Spatial Audio Encoder Architecture}
\samethod's architecture is inspired from Audio Spectrogram Transformer~\citep{ast} with a causal self-attention backbone. The input to the model is the multi-channel complex spectrogram $\mathbf{X} \in \mathbb{C}^{B \times F \times T \times C}$, where $B$ is the batch size. We interleave the real and imaginary parts of the spectrogram along the channel dimension to form a tensor of shape $[B, F, T, 2C=10]$. 

To encode the audio, we first reshape the input by flattening the frequency and channel dimensions, yielding a representation $\mathbf{X}' \in \mathbb{R}^{B \times T \times (F \cdot 2C)}$. We then extract temporal patches using a 1D convolution over the time dimension with a kernel size and stride equal to the patch temporal resolution $P=10$, which corresponds to 100 ms. This projection maps the high-dimensional spectrogram into a sequence of continuous embeddings $\mathbf{E} \in \mathbb{R}^{B \times L \times D}$, where $L = 300$ is the sequence length and $D = 512$ is the hidden dimension. Following patch extraction, we add learnable 1D positional embeddings $\mathbf{P} \in \mathbb{R}^{1 \times L \times D}$ to the continuous embeddings $\mathbf{E}$ to retain temporal order. The resulting sequence $\mathbf{H}^{(0)}$ is then passed through a stack of $N$ transformer blocks. We enforce a strict causal self-attention mask so that the spatial encoder can support future streaming use; the current end-to-end model is not streaming because its frozen monaural audio encoder is non-causal.

Because our context window is 30 seconds, audio inputs shorter than this duration are padded with zeros. We produce an attention mask for the encoder of shape $[B, L]$ that indicates which frames in the sequence of length $L$ are valid and which are padding. This mask is utilized as an attention mask within the transformer layers. The final output sequence is denoted as $\mathbf{H} \in \mathbb{R}^{B \times L \times D}$.

\subsubsection{Encoder Training Objectives}
\samethod{} is trained jointly on frame-level wearer Voice Activity Detection (VAD) and Sound Event Localization and Detection (SELD), using the coordinate system specified in Appendix~\ref{appendix:cord_sys}. Both heads output frame-wise predictions where each frame is 100 ms, resulting in $L = 300$ frames. The VAD head predicts wearer-speech activity with binary cross-entropy. For the SELD head, we use the Multi-ACCDOA format \citep{shimada2022multi} with two classes (speech and non-speech) and three tracks per class to regress Cartesian directions. To address the track permutation problem, we employ the Location-based Training (LBT) loss \citep{scheibler2022sorting}. The total loss is the addition of the VAD and SELD losses with equal weights to jointly train the encoder; afterward, both task-specific heads are discarded and the frozen encoder supplies spatial embeddings to \method{}. Appendix~\ref{appendix:foacoder_pretraining_objectives} provides the complete target construction and objective definitions.

\subsection{Large-Scale Synthetic Spatial Audio Generation}
\label{subsec:synthetic_data}

To train \samethod{} across diverse arrays with varying microphone counts and geometries, we synthesized $\sim$2M multi-source 3D acoustic scenes. This simulation pipeline captures diverse overlapping sound sources and precise ground-truth labels.

\subsubsection{Array Transfer Function Simulations}
\label{subsubsec:atf_simulation}

We use Computer-Aided Design (CAD) models of two smart glasses featuring 9 and 15 microphones, fitted onto a head and torso (HAT) model. Appendix~\ref{appendix:glasses_arrays} illustrates both microphone-array geometries and the Design~1 glasses on the HAT model. Using COMSOL Multiphysics \citep{comsol}, we simulate anechoic Array Transfer Functions (ATFs) to model wave propagation, scattering, and head occlusion (details in Appendix~\ref{appendix:atf_simulation_details}). These array transfer functions are used for both constructing the Ambisonic Signal Matching beamformer weights (Section~\ref{subsubsec:pre_processing}) and for the room acoustic simulations described in Section~\ref{subsec:rir_generation}.

\subsubsection{Array Selection}
\label{subsubsec:array_selection}

Not all microphone combinations can accurately reconstruct the FOA signal (e.g., co-planar configurations cannot reconstruct the Z channel). Through a beamforming simulation study, we select 1,211 unique 4- to 9-microphone array configurations that can reconstruct the FOA audio sufficiently. We randomly sample one of these array configurations per generated spatial audio scene. The details of the simulation methodology are provided in Appendix \ref{appendix:asm_simulation}, and the array selection criteria are outlined in Appendix \ref{appendix:beamforming_artifacts}.

\subsubsection{Room Impulse Response (RIR) Generation}
\label{subsec:rir_generation}

To model realistic reverberation, we use an image-source room acoustics simulator \citep{allen1979image}. We generate random shoebox-shaped rooms with varied dimensions and acoustic materials. We place the wearer and two external sound sources randomly inside the room. Crucially, we integrate the anechoic ATFs into these simulations to account for head and device scattering. This yields $\sim$200k/20k/20k multi-source Room Impulse Responses (RIRs) for training, validation, and testing (Appendix \ref{appendix:rir_simulation_details}).

\subsubsection{Source Audio Sampling and Mixing}
\label{subsec:source_audio_sampling_and_mixing}

We sample monophonic speech from LibriTTS~\citep{libritts} and environmental sounds from AudioSet~\citep{audioset} and Freesound~\citep{fonseca2017freesound}. Each scene contains a randomized mix of wearer speech, bystander speech, and non-speech sounds. We convolve these signals with the simulated RIRs and additively mix them with randomized gains and delays. We intentionally create dense, overlapping ``cocktail party'' scenarios rather than structured turn-taking. This produces $\sim$2M/150k/150k 30-second spatial audio examples with rich 3D spatial metadata. (More details in Appendix \ref{appendix:mixing_parameters}).

\subsection{\methodfull{}}
\label{subsec:method_architecture}

We construct \method{} by augmenting \basemodel{} with a second audio pathway for the continuous embeddings produced by \samethod{}. For every 30-second scene, the model conditions jointly on 750 embeddings from the base model's native audio encoder, 300 embeddings from the spatial encoder, and the text instruction. Both audio representations are inserted directly into the language model's input embedding sequence. Finally, the model is trained to generate structured outputs describing the acoustic scene, as detailed in Section~\ref{subsubsec:spatial_audio_sft}.

\subsubsection{Dual Audio Pathway}
The two pathways operate at complementary temporal resolutions. In the \textbf{native audio pathway}, the mono signal is converted to Mel features and passed through the pretrained Gemma audio tower and audio projection. Its output is padded or truncated to 750 embeddings, corresponding to one embedding every 40~ms. In the \textbf{spatial audio pathway}, the FOA-plus-reference-channel representation is passed through the frozen \samethod{} encoder described in Section~\ref{subsec:sa_encoder}. It produces 300 512-dimensional embeddings, corresponding to one embedding every 100~ms, together with a validity mask for recordings shorter than 30 seconds. Thus, the native pathway supplies high-resolution acoustic and semantic evidence, while the spatial pathway provides the multichannel cues learned under the SELD and wearer-VAD training objectives.

\subsubsection{Learned Spatial Projector and LLM Injection}
The spatial projector follows Gemma's multimodal embedding design rather than a plain MLP. For each spatial embedding $\mathbf{h}_t\in\mathbb{R}^{512}$, the projector applies a learned RMS normalization, a bias-free $512\rightarrow D_{\mathrm{LLM}}$ linear layer (where $D_{\mathrm{LLM}}$ denotes the hidden dimension of the language model), a GELU activation, a second bias-free $D_{\mathrm{LLM}}\rightarrow D_{\mathrm{LLM}}$ linear layer, and a post-projection RMS normalization. The spatial block always occupies 300 input tokens. For recordings shorter than 30 seconds, the frame-validity mask identifies padded \samethod{} outputs but does not remove their corresponding placeholders from LLM self-attention. We therefore replace each invalid frame after projection with a shared learned padding embedding passed through the same projection path, explicitly distinguishing absent spatial observations from valid projected features while preserving the fixed-length input layout.

We extend the base tokenizer with three new structural tokens: \texttt{<start\_of\_spatial\_audio>}, \texttt{<spatial\_audio\_placeholder>}, and \texttt{<end\_of\_spatial\_audio>}. The prompt contains 300 spatial placeholder tokens between the two boundaries. After text embedding lookup, these placeholders are overwritten by the projected \samethod{} outputs. Likewise, the 750 native-audio placeholder tokens, each represented by \texttt{<|audio|>}, are overwritten by the native audio embeddings. The resulting sequence of text, native-audio, and spatial-audio embeddings is processed jointly by the language model's self-attention.

\subsubsection{Spatial Audio Supervised Fine-Tuning}
\label{subsubsec:spatial_audio_sft}

The \method{} model is trained via supervised fine-tuning (SFT) on questions and responses spanning six task types, returning a structured JSON list of sound sources. Each source entry specifies the sound type, start and end times, and, for external (non-wearer) sources, the wearer-relative location given by numerical azimuth, elevation, and distance values alongside categorical direction and distance descriptors (bins defined in Appendices~\ref{sec:full_metrics_suite} and~\ref{sec:direction_bins}). Speech sources also include a transcript, whereas wearer speech contains a transcript and timestamps but no spatial attributes because it originates on the wearer.

We formulate six distinct tasks. \textit{Task 1: Holistic Sound Localization} asks the model to localize and predict the timestamps of all external sound events. \textit{Task 2: Selective Sound Localization} asks the model to return the locations and time intervals of all instances of a queried external non-speech class. \textit{Task 3: Holistic Captioning} requires the model to detect, localize, and transcribe all wearer speech, bystander (non-wearer) speech, and external non-speech sounds. \textit{Task 4: Role-Aware Transcription} requires the model to transcribe all wearer and bystander speech while providing the speaker roles, time boundaries, and spatial locations for bystanders. \textit{Task 5: Targeted Wearer Transcription} instructs the model to transcribe only the speech produced by the wearer. \textit{Task 6: Targeted External Transcription} queries the model to transcribe only bystander speech arriving from a specific categorical direction.

For the two selective tasks (Task 2 and Task 6), we also intentionally generate negative examples by querying the model for an absent sound class or an unoccupied direction. This trains the model to recognize missing targets and output an empty response. We refer the reader to Appendix~\ref{sec:qa_examples} for further details regarding Q\&A generation.

\subsubsection{Spatio-Temporal Weighted Cross-Entropy}
The structured targets contain long speech transcripts but comparatively few spatial coordinates and sound-event time boundaries. Uniform next-token cross-entropy can therefore allocate most of its aggregate training signal to transcription. We address this imbalance by assigning larger loss weights to target tokens that encode spatial coordinates and sound-event boundaries.

For each target response, we locate the character spans of the numeric values corresponding to the azimuth, elevation, and distance, as well as the start and end times of the sound events, identifying every target token whose character interval overlaps one of these values. Let $\mathcal{I}$ be the supervised target-token positions, and let the disjoint subsets $\mathcal{I}_{\mathrm{spatial}}$ and $\mathcal{I}_{\mathrm{temporal}}$ contain the positions aligned to these spatial and temporal values, respectively. Letting $\ell_t=-\log p_\theta(y_t\mid y_{<t},\mathbf{x})$ denote the standard autoregressive next-token prediction loss, our weighted objective is
\begin{equation}
\mathcal{L}_{\text{ST-WCE}}
=
\frac{\sum_{t\in\mathcal{I}} w_t\ell_t}
{\sum_{t\in\mathcal{I}} w_t},
\qquad
w_t = \begin{cases}
\alpha_{\mathrm{spatial}}, & t\in\mathcal{I}_{\mathrm{spatial}},\\
\alpha_{\mathrm{temporal}}, & t\in\mathcal{I}_{\mathrm{temporal}},\\
1, & \text{otherwise}.
\end{cases}
\label{eq:spatiotemporal_weighted_ce}
\end{equation}
Prompt, multimodal-placeholder, and padding positions are excluded from $\mathcal{I}$. Because the two position sets are derived from distinct JSON keys, they do not overlap. We use $(\alpha_{\mathrm{spatial}},\alpha_{\mathrm{temporal}})=(5,5)$ for both Stage~1 projector alignment and Stage~2 holistic SFT. We selected these weights empirically using validation-set evaluation loss after comparing $(1,1)$, $(1,5)$, $(5,1)$, $(5,5)$, and $(10,10)$ while holding the data, curriculum, and all other hyperparameters fixed. Normalizing by the sum of valid weights keeps the objective scale comparable across examples with different source counts and transcript lengths.

\section{Evaluation}

We evaluate \method{} end-to-end on six spatial-audio understanding tasks. We also evaluate \samethod{} independently on SELD and wearer-VAD objectives; Appendix~\ref{appendix:foacoder_evaluation} provides the encoder evaluation protocol, baselines, metrics, and complete results.

\label{sec:sealm_eval}
\subsection{\method{} Evaluation Set}
We evaluate \method{} on a held-out synthetic evaluation set generated using the simulation procedures from Section~\ref{subsec:synthetic_data}. This produces varying acoustic complexities from complete silence to mixtures of up to three sound sources. These sources might be completely non-overlapping, or they might partially or completely overlap (see Appendix~\ref{appendix:mixing_parameters} for details).

We use only held-out monophonic test audio and a disjoint test set of simulated RIRs, so neither source waveforms nor acoustic responses overlap with training. The evaluation set contains 136,378 scene--question pairs across all six tasks: 26,985 for Task~1, 17,247 for Task~2, 33,977 for Task~3, 25,876 for Task~4, 17,372 for Task~5, and 14,921 for Task~6. Each pair consists of one audio scene and one question for a task supported by that scene; a scene does not necessarily contribute an example to every task. As in training, we select one task uniformly at random from the valid tasks supported by a scene; Tasks~2 and~6 use an absent class or direction with probability 0.3 to evaluate negative-query rejection. Prompt construction and structured targets otherwise follow Section~\ref{subsubsec:spatial_audio_sft} exactly.

\subsection{\method{} Evaluation Metrics}
\label{sec:sealm_eval_metrics}

To evaluate each model response, we first verify whether the output adheres to the required structured JSON format. For validly formatted predictions, computing most metrics requires identifying which predicted source corresponds to which ground-truth source, as their counts may differ. We match external predicted and ground-truth sources using the Hungarian algorithm based on their spatial and temporal agreement, whereas wearer speech is matched directly since its ego-centric origin lacks spatial coordinates. Across the six tasks, we report format adherence, hallucinated and missing sources, spatial and semantic accuracy, temporal localization, speech WER, and wearer- or query-detection metrics as applicable. Because spatial, temporal, source-type, and transcription errors are computed only over matched pairs of predicted and ground-truth sources, they must be interpreted alongside the hallucination and missing-source rates, which measure unmatched predictions and undetected ground-truth sources. Appendix~\ref{sec:full_metrics_suite} provides full details on how we match predicted and ground-truth sources and how each metric is computed. Additionally, Appendix~\ref{appendix:evaluation_set_cohorts} details our cohort analysis, which breaks down model performance across dataset subsets grouped by source count, speaker count, source overlap, and microphone count. Table~\ref{tab:evaluation_metrics_main} summarizes the metrics applicable to each task.

\begin{table*}[h]
\centering
\scriptsize
\caption{Evaluation metrics by task: T1, Holistic Sound Localization; T2, Selective Sound Localization; T3, Holistic Captioning; T4, Role-Aware Transcription; T5, Targeted Wearer Speech Transcription; and T6, Targeted External Speech Transcription. Accuracy, precision, recall, and F1 are reported for both wearer and query detection where applicable.}
\label{tab:evaluation_metrics_main}
\renewcommand{\cmark}{\textcolor{green!60!black}{\ding{51}}}
\renewcommand{\xmark}{\textcolor{red!75!black}{\ding{55}}}
\setlength{\tabcolsep}{2.5pt}
\begin{tabular}{@{}p{0.25\textwidth}p{0.48\textwidth}*{6}{c}@{}}
\toprule
\textbf{Metric} & \textbf{Definition} & \textbf{T1} & \textbf{T2} & \textbf{T3} & \textbf{T4} & \textbf{T5} & \textbf{T6} \\
\midrule
Format adherence & Percentage of outputs that parse and satisfy the required schema. & \cmark & \cmark & \cmark & \cmark & \cmark & \cmark \\
\hdashline
External-source hallucination and missing-source rates & Percentage of validly parsed examples with at least one unmatched predicted or ground-truth external source. & \cmark & \cmark & \cmark & \cmark & \xmark & \cmark \\
\hdashline
Source-type accuracy & Exact agreement between the sound descriptors of matched predicted and ground-truth external sources after text normalization and mapping synonyms to canonical labels. & \cmark & \xmark & \cmark & \xmark & \xmark & \xmark \\
\hdashline
Azimuth, elevation, and distance MAE; percentage distance error & Spatial localization errors over matched predicted and ground-truth external sources. & \cmark & \cmark & \cmark & \cmark & \xmark & \cmark \\
\hdashline
Location- and distance-bin accuracy & Exact agreement between the categorical spatial bins of matched predicted and ground-truth external sources. & \cmark & \cmark & \cmark & \cmark & \xmark & \cmark \\
\hdashline
Start- and end-time MAE; temporal IoU & Temporal boundary error and interval overlap over matched predicted and ground-truth sources. & \cmark & \cmark & \cmark & \cmark & \cmark & \cmark \\
\hdashline
Speech transcription WER & Capped WER on normalized transcripts from matched speech sources. & \xmark & \xmark & \cmark & \cmark & \cmark & \cmark \\
\hdashline
Wearer detection (accuracy, precision, recall, F1) & Example-level detection of whether wearer speech is present. & \xmark & \xmark & \cmark & \cmark & \cmark & \xmark \\
\hdashline
Query detection (accuracy, precision, recall, F1) & Source-level detection of the queried sound class or direction. & \xmark & \cmark & \xmark & \xmark & \xmark & \cmark \\
\bottomrule
\end{tabular}
\vspace{-2em}
\end{table*}

\subsection{Baselines}
We compare \method{} against three baselines. \textbf{Vanilla \basemodel{}} is the pretrained model used without task-specific adaptation. \textbf{Fine-tuned \basemodel{}} is a controlled monaural baseline that retains only the model's native audio pathway and is fine-tuned on our synthetic data. \textbf{SELDNet $+$ \basemodel{}} is inspired by the frozen-SELDNet baseline in Sci-Phi~\citep{sciphi2025}: FOA log-mel and intensity-vector features are passed through a SELDNet trunk initialized from the \href{https://github.com/partha2409/DCASE2024_seld_baseline/}{DCASE 2024 baseline}. To ensure a controlled comparison, we keep the LLM backbone, native audio encoder, synthetic SFT data, and two-stage training curriculum (Section~\ref{sec:experiments}) identical to \method{}, changing only the spatial encoder.

\section{Experiments}
\label{sec:experiments}
\noindent\textbf{Training Setup for \samethod.}
We train \samethod{} for 2,000,000 steps with a batch size of 128 under joint VAD and SELD supervision on 8$\times$NVIDIA H100 GPUs; Appendix~\ref{appendix:foacoder_training_setup} provides the complete optimizer and learning-rate configuration.

\noindent\textbf{Training Setup for \method.}
We train \method{} only on synthetic data. The two stages of training follow a curriculum training scheme, and we start with an easier training problem for Stage~1. Both stages use 8$\times$H100 GPUs, a 2048-token context length, bfloat16 precision, the AdamW optimizer, and a cosine decay learning rate schedule with 5\% warmup. Stage~1 optimizes only the spatial projector for 3 epochs on Task~1, where examples are limited to ones that have a single sound source ($5\times10^{-4}$ learning rate; batch 4/GPU with two accumulation steps). Stage~2 warm-starts from Stage~1 and trains for 3 epochs on all six tasks with one to four sound sources ($1\times10^{-4}$ learning rate; batch 2/GPU with four accumulation steps), optimizing the projector and rank-128 LLM LoRA adapters while freezing \samethod{}, the native audio tower, and base LLM. Both stages have an effective global batch size of 64. Appendices~\ref{appendix:training_stages} and~\ref{appendix:curse_curriculum} provide full curriculum details and an ablation showing that inserting a third intermediate training stage degrades performance.

Both fine-tuned baselines follow the same two-stage curriculum and synthetic training data as \method{}: single-source Task~1 examples in Stage~1, followed by all six tasks with the full source-count mixture in Stage~2. \textbf{SELDNet $+$ \basemodel{}} uses the same trainable and frozen component split as \method{}, optimizing only the spatial projector in Stage~1 and the projector plus LLM LoRA adapters in Stage~2, while keeping the SELDNet spatial encoder frozen throughout. \textbf{Fine-tuned \basemodel{}} retains only the native monaural audio pathway, ensuring that its comparison with \method{} isolates the contribution of spatial audio conditioning.

\section{Results}
\label{sec:results}

We focus the main-paper analysis on the end-to-end \method{} results. Independent \samethod{} results and their full discussion are reported in Appendix~\ref{appendix:foacoder_evaluation}.

At the encoder level, the selected 30-second \samethod{} reduces angular error from $44.01^\circ$ to $15.06^\circ$ and raises localization recall from $32.29\%$ to $86.50\%$ relative to the encoder-only SELDNet reference, while achieving $98.53\%$ wearer-VAD F1. This indicates that the frozen spatial pathway preserves both external-source geometry and egocentric speech activity before language-model integration.

\subsection{\method{} Results}
\label{sec:results}

\begin{table*}[t]
\centering
\small
\caption{Task-wise results on the evaluation set. Best values are \textbf{bold}; second-best values are \underline{underlined}.\footnotesize \textit{ Note:} Azim., Elev., Dist., Type acc., and tIoU denote azimuth MAE, elevation MAE, distance MAE, source-type accuracy, and temporal intersection over union, respectively. Halluc. and Missing denote the external-source hallucination and missing-source rates. $^{\dagger}$ Query F1 is the F1 detection score for the queried source type in Task~2 or the queried direction in Task~6.}
\label{tab:main_results_compact}
\vspace{-1em}
\resizebox{\linewidth}{!}{
\begin{tabular}{@{}l*{8}{c}@{}}
\toprule
\multicolumn{9}{c}{\textbf{(a) Spatial and event understanding (Tasks 1--2).}} \\ \midrule
& \multicolumn{5}{c}{\textbf{Task 1: Holistic Sound Localization}} & \multicolumn{3}{c}{\textbf{Task 2: Selective Sound Localization}} \\
\cmidrule(lr){2-6} \cmidrule(lr){7-9}
\textbf{Model} & \textbf{Azim.} $\downarrow$ & \textbf{Elev.} $\downarrow$ & \textbf{Dist.} $\downarrow$ & \textbf{Type acc.} $\uparrow$ & \textbf{tIoU} $\uparrow$ & \textbf{Azim.} $\downarrow$ & \textbf{Elev.} $\downarrow$ & \textbf{Query F1}$^{\dagger}$ $\uparrow$ \\
& ($^\circ$) & ($^\circ$) & (m) & (\%) & & ($^\circ$) & ($^\circ$) & (\%) \\
\midrule
\basemodel{} (vanilla) & \underline{67.9} & 27.7 & 1.88 & 0.9 & 0.37 & \underline{48.0} & \underline{26.4} & 10.9 \\
\basemodel{} (fine-tuned) & 87.5 & 27.7 & 1.28 & \underline{62.2} & \underline{0.91} & 88.2 & 28.3 & \underline{78.0} \\
SELDNet $+$ \basemodel{} & 80.6 & \underline{27.4} & \underline{1.26} & 55.6 & 0.61 & 80.2 & 27.4 & 76.2 \\
\rowcolor{methodrow}
\method{} \textit{(ours)} & \textbf{14.9} & \textbf{12.8} & \textbf{1.07} & \textbf{63.7} & \textbf{0.92} & \textbf{17.5} & \textbf{14.8} & \textbf{94.9} \\
\bottomrule
\end{tabular}}

\resizebox{\linewidth}{!}{
\begin{tabular}{@{}l*{11}{c}@{}}
\toprule
\multicolumn{12}{c}{\textbf{(b) Transcription and source detection (Tasks 3--6).}} \\ \midrule
& \multicolumn{4}{c}{\textbf{Task 3: Holistic Captioning}} & \multicolumn{3}{c}{\textbf{Task 4: Role-Aware Transcription}} & \multicolumn{2}{c}{\textbf{Task 5: Targeted Wearer Transcription}} & \multicolumn{2}{c}{\textbf{Task 6: Targeted External Transcription}} \\
\cmidrule(lr){2-5} \cmidrule(lr){6-8} \cmidrule(lr){9-10} \cmidrule(lr){11-12}
\textbf{Model} & \textbf{Azim.} $\downarrow$ & \textbf{Elev.} $\downarrow$ & \textbf{Dist.} $\downarrow$ & \textbf{WER} $\downarrow$ & \textbf{WER} $\downarrow$ & \textbf{Halluc.} $\downarrow$ & \textbf{Missing} $\downarrow$ & \textbf{WER} $\downarrow$ & \textbf{Wearer F1} $\uparrow$ & \textbf{WER} $\downarrow$ & \textbf{Query F1}$^{\dagger}$ $\uparrow$ \\
& ($^\circ$) & ($^\circ$) & (m) & (\%) & (\%) & (\%) & (\%) & (\%) & (\%) & (\%) & (\%) \\
\midrule
\basemodel{} (vanilla) & \underline{65.3} & 27.6 & 1.71 & 24.5 & 27.7 & 15.8 & 45.0 & 38.5 & \underline{87.5} & 49.0 & 13.2 \\
\basemodel{} (fine-tuned) & 86.7 & 27.6 & 1.25 & 21.6 & 21.8 & 15.6 & \underline{3.4} & 8.9 & \textbf{100.0} & 38.1 & 83.4 \\
SELDNet $+$ \basemodel{} & 80.3 & \underline{27.4} & \underline{1.22} & \underline{19.3} & \textbf{17.7} & \underline{10.9} & 16.1 & \underline{6.5} & \textbf{100.0} & \underline{36.2} & \underline{86.9} \\
\rowcolor{methodrow}
\method{} \textit{(ours)} & \textbf{14.2} & \textbf{12.8} & \textbf{1.06} & \textbf{17.8} & \underline{17.8} & \textbf{4.5} & \textbf{1.8} & \textbf{5.8} & \textbf{100.0} & \textbf{32.5} & \textbf{94.1} \\
\bottomrule
\end{tabular}}
\vspace{-1.5em}
\end{table*}

Table~\ref{tab:main_results_compact} compares \method{} with three baseline models on the evaluation set. We computed an extensive set of metrics that were explained in Section~\ref{sec:sealm_eval_metrics}. In this section, we report a chosen subset of these metrics. However, the full results can be found in Appendix~\ref{appendix:evaluation_set_complete}. Appendix~\ref{appendix:evaluation_set_cohorts} presents the results of the cohort analysis, where \method{} performance is stratified by sound source number, speech overlap, and microphone count.

\noindent\textbf{Evaluation-Set Results.}
\method{} performs best on nearly every metric summarized in Table~\ref{tab:main_results_compact}; the only exception is Role-Aware Transcription WER, for which SELDNet $+$ \basemodel{} is marginally better. Compared with this  baseline, \method{} substantially reduces localization error and predicts more precise temporal event boundaries. It also reduces external-source hallucinations and omissions while improving transcription accuracy for wearer and external speech.

\noindent\textbf{Cohort Analysis.}
Performance degrades as source count and overlap increase, with the largest effect on external-speech transcription; wearer-speech transcription remains comparatively robust. Across the evaluated 4--9-microphone subsets, performance remains stable, although the fixed glasses geometries limit conclusions about arbitrary arrays or the general effect of adding microphones. Appendix~\ref{appendix:evaluation_set_cohorts} provides the complete cohort analysis.

\section{Conclusion, Limitations \& Future Work}
\label{sec:conclusion}

\noindent\textbf{Conclusion.}
We presented \method{}, an ego-centric spatial audio-language model that jointly reasons about who and what is audible, where and when events occur, and what is spoken. We demonstrate a spatial audio data synthesis pipeline to generate non-ideal FOA spatial audio, alongside \samethod{}, an encoder capable of extracting spatial embeddings from this data. The model's dual audio pathway incorporates these embeddings into an LLM by combining them with semantic audio features, while spatio-temporal weighted cross-entropy preserves supervision for sparse location and time values. Through a curriculum learning approach, we show how the model learns to understand spatial audio. On our evaluation set, \method{} outperforms monaural and SELDNet-based baselines on most reported metrics, maintaining reliable wearer speech understanding under dense source overlap. It also demonstrates robustness without array-specific retraining across 1,211 evaluated glasses arrays with 4 to 9 microphones and varied microphone positions.

\noindent\textbf{Limitations and Future Work.}
Our study focuses on simulated, stationary scenes and 1,211 configurations derived from two smart-glasses geometries, enabling controlled evaluation across a broad range of microphone counts and positions. Extending this evaluation to real and dynamic scenes, held-out array topologies, and other device form factors, together with exhaustive component, uncertainty, and streaming studies, was beyond the scope of this work due to complexities in simulating moving sources, and post processing existing datasets while maintaining data quality and realism, and is left to future research. We will also investigate improving external-speech transcription under dense source overlap by incorporating finetuning the mono audio encoder on cocktail scenes.

\section{Acknowledgments}
We thank Hakan Erdogan for his valuable feedback on the manuscript and his contributions to the geometrical acoustics simulator. We also thank Jeff Tsai for configuring and executing the COMSOL array transfer function simulations, and Yvette Su for conducting the post-processing of the simulation data.

\section{Ethics Statement}
The authors acknowledge and adhere to the ICLR Code of Ethics. This study introduces no new human-subject data collection; it uses synthetic acoustic scenes and established speech and environmental-audio datasets. Spatial localization, role attribution, and transcription can support accessibility and embodied interaction, but can also enable privacy-invasive monitoring or inference about people nearby. Deployment should therefore require appropriate consent, data minimization, secure audio handling, and human oversight, especially in private or safety-critical settings. Moreover, biases in the source corpora, languages, acoustic environments, and simulated device geometries may yield uneven performance across speakers and deployment conditions; evaluation on representative populations and environments is necessary before consequential use.

\section{Reproducibility Statement}
We document the \samethod{} and \method{} architectures, objectives, structured output format, and spatio-temporal loss in Sections~\ref{subsec:sa_encoder}--\ref{subsec:method_architecture}; synthetic data generation and task construction in Sections~\ref{subsec:synthetic_data} and~\ref{subsubsec:spatial_audio_sft}; and training and evaluation procedures in Sections~\ref{sec:experiments} and~\ref{sec:sealm_eval_metrics}. The appendix specifies the independent encoder evaluation (Appendix~\ref{appendix:foacoder_evaluation}), training stages (Appendix~\ref{appendix:training_stages}), array and room simulation (Appendices~\ref{appendix:atf_simulation_details} and~\ref{appendix:rir_simulation_details}), source mixing (Appendix~\ref{appendix:mixing_parameters}), ASM beamforming (Appendix~\ref{appendix:asm_beamforming_details}), task generation and examples (Appendix~\ref{sec:qa_examples}), and cohort and curriculum analyses (Appendices~\ref{appendix:evaluation_set_cohorts} and~\ref{appendix:curse_curriculum}). Appendix~\ref{appendix:ai_usage} provides the AI-usage disclosure.

\bibliography{iclr2027_conference}
\bibliographystyle{iclr2027_conference}

\newpage

\appendix
\section*{Appendix Contents}

\begingroup
\small
\noindent\hyperref[appendix:cord_sys]{\textbf{Appendix~\ref*{appendix:cord_sys}:} Coordinate System}\dotfill\hyperref[appendix:cord_sys]{\pageref*{appendix:cord_sys}}\par
\noindent\hyperref[appendix:asm_beamforming_details]{\textbf{Appendix~\ref*{appendix:asm_beamforming_details}:} Ambisonic Signal Matching Beamforming Details}\dotfill\hyperref[appendix:asm_beamforming_details]{\pageref*{appendix:asm_beamforming_details}}\par
\noindent\hyperref[appendix:beamforming_artifacts]{\textbf{Appendix~\ref*{appendix:beamforming_artifacts}:} Beamforming Artifacts and Array Selection}\dotfill\hyperref[appendix:beamforming_artifacts]{\pageref*{appendix:beamforming_artifacts}}\par
\noindent\hyperref[appendix:foacoder_evaluation]{\textbf{Appendix~\ref*{appendix:foacoder_evaluation}:} \samethod{} Training and Evaluation}\dotfill\hyperref[appendix:foacoder_evaluation]{\pageref*{appendix:foacoder_evaluation}}\par
\noindent\hyperref[appendix:glasses_arrays]{\textbf{Appendix~\ref*{appendix:glasses_arrays}:} Glasses Microphone Arrays}\dotfill\hyperref[appendix:glasses_arrays]{\pageref*{appendix:glasses_arrays}}\par
\noindent\hyperref[appendix:atf_simulation_details]{\textbf{Appendix~\ref*{appendix:atf_simulation_details}:} Array Transfer Function Simulation Details}\dotfill\hyperref[appendix:atf_simulation_details]{\pageref*{appendix:atf_simulation_details}}\par
\noindent\hyperref[appendix:asm_simulation]{\textbf{Appendix~\ref*{appendix:asm_simulation}:} Ambisonic Signal Matching Simulation Study}\dotfill\hyperref[appendix:asm_simulation]{\pageref*{appendix:asm_simulation}}\par
\noindent\hyperref[appendix:rir_simulation_details]{\textbf{Appendix~\ref*{appendix:rir_simulation_details}:} Room Impulse Response Simulation Details}\dotfill\hyperref[appendix:rir_simulation_details]{\pageref*{appendix:rir_simulation_details}}\par
\noindent\hyperref[appendix:mixing_parameters]{\textbf{Appendix~\ref*{appendix:mixing_parameters}:} Mixing Parameters}\dotfill\hyperref[appendix:mixing_parameters]{\pageref*{appendix:mixing_parameters}}\par
\noindent\hyperref[sec:qa_examples]{\textbf{Appendix~\ref*{sec:qa_examples}:} Question and Answer Generation Details and Examples}\dotfill\hyperref[sec:qa_examples]{\pageref*{sec:qa_examples}}\par
\noindent\hyperref[sec:full_metrics_suite]{\textbf{Appendix~\ref*{sec:full_metrics_suite}:} \method{} Evaluation Protocol and Metrics}\dotfill\hyperref[sec:full_metrics_suite]{\pageref*{sec:full_metrics_suite}}\par
\noindent\hyperref[appendix:evaluation_set_cohorts]{\textbf{Appendix~\ref*{appendix:evaluation_set_cohorts}:} Cohort Analysis}\dotfill\hyperref[appendix:evaluation_set_cohorts]{\pageref*{appendix:evaluation_set_cohorts}}\par
\noindent\hyperref[sec:direction_bins]{\textbf{Appendix~\ref*{sec:direction_bins}:} Categorical Direction Bins}\dotfill\hyperref[sec:direction_bins]{\pageref*{sec:direction_bins}}\par
\noindent\hyperref[appendix:training_stages]{\textbf{Appendix~\ref*{appendix:training_stages}:} Training Stages for \method{}}\dotfill\hyperref[appendix:training_stages]{\pageref*{appendix:training_stages}}\par
\noindent\hyperref[appendix:curse_curriculum]{\textbf{Appendix~\ref*{appendix:curse_curriculum}:} Curriculum Ablation Studies}\dotfill\hyperref[appendix:curse_curriculum]{\pageref*{appendix:curse_curriculum}}\par
\noindent\hyperref[appendix:evaluation_set_complete]{\textbf{Appendix~\ref*{appendix:evaluation_set_complete}:} Extended \method{} Evaluation Results}\dotfill\hyperref[appendix:evaluation_set_complete]{\pageref*{appendix:evaluation_set_complete}}\par
\noindent\hyperref[appendix:ai_usage]{\textbf{Appendix~\ref*{appendix:ai_usage}:} AI Usage}\dotfill\hyperref[appendix:ai_usage]{\pageref*{appendix:ai_usage}}\par
\endgroup

\section{Coordinate System}
\label{appendix:cord_sys}
\begin{figure}[h]
    \centering
    \includegraphics[width=0.45\textwidth]{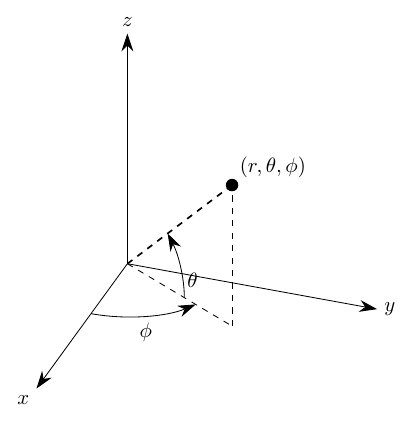}
    \caption{Spherical coordinate system showing distance $r$, elevation angle $\theta$ (measured from the $xy$-plane), and azimuth angle $\phi$.}
    \label{fig:coord_sys}
\end{figure}

Figure~\ref{fig:coord_sys} illustrates the spherical coordinate system used in this work. We orient the wearer of the device (such as smart glasses) such that they are facing the positive $x$-direction; from the perspective of the wearer, $+x$-axis points forward, $+y$-axis points left, and $+z$-axis points up. We define $\theta$ as the elevation angle measured upwards from the $xy$-plane towards the point. The azimuth angle $\phi$ is measured from the positive $x$-axis in the $xy$-plane.

\section{Ambisonic Signal Matching Beamforming Details}
\label{appendix:asm_beamforming_details}

We compute the short-time Fourier Transform (STFT) of the multichannel audio signal before applying the beamformer. For the STFT, we use a Hann window with a window and FFT size of 512 samples and a hop size of 256.

We follow \citep{gayer2024ambisonics} to compute the beamforming coefficients and apply them to the input signal. Let $\mathbf{x}(f) \in \mathbb{C}^{N_{\text{mics}}}$ be the frequency-domain multichannel signal at a single STFT time-frame at frequency $f$, where $N_{\text{mics}}$ is the number of microphones. We denote:

\begin{equation*}
\hat{\mathbf{a}}(f) = [\hat{a}_{00}(f), \hat{a}_{1-1}(f), \hat{a}_{10}(f), \hat{a}_{11}(f)]^T \in \mathbb{C}^{(N_a + 1)^2} 
\end{equation*}

\noindent as the full estimated FOA signal vector (the output of the ASM beamformer), where $N_a=1$ is the Ambisonic order. We want to compute the beamforming weights $\mathbf{w}_{nm}(f) \in \mathbb{C}^{N_{\text{mics}}}$ for each frequency bin $f$, Ambisonic order $n$, and degree $m$. These weights are applied to the microphone signals to obtain the estimated FOA scalar component $\hat{a}_{nm}(f)$ as:

\begin{equation}
\hat{a}_{nm}(f) = \mathbf{w}_{nm}(f)^H \mathbf{x}(f)
\label{eq:beamformer_apply}
\end{equation}

\noindent Provided with simulated far-field-to-microphone transfer functions (as described in Section \ref{subsubsec:atf_simulation}) $H_{ff,i}(\theta_q, \phi_q, f) \in \mathbb{C}$ for each microphone $i$ and for a grid of points $(\theta_q, \phi_q) \ \forall \ 1 \leq q \leq Q$, we construct the steering matrix $\mathbf{V}(f) \in \mathbb{C}^{N_{\text{mics}} \times Q}$:

\begin{equation}
[\mathbf{V}(f)]_{i,q} = H_{ff,i}(\theta_q, \phi_q, f)
\label{eq:steering_matrix}
\end{equation}

\noindent The beamforming weights are then computed as:

\begin{equation}
  \mathbf{w}_{nm}(f) = \left( \mathbf{V}(f)\mathbf{V}(f)^H + \lambda \mathbf{I} \right)^{-1} \mathbf{V}(f)\mathbf{y}_{nm}
  \label{eq:beamformer_weights}
\end{equation}

\noindent where $\mathbf{w}_{nm}(f) \in \mathbb{C}^{N_{\text{mics}}}$ represents the beamforming weights, and $\lambda$ is the Tikhonov regularization parameter. We pick $\lambda=10^{-3}$ empirically. The vector $\mathbf{y}_{nm} \in \mathbb{R}^{Q}$ contains the spherical harmonic functions of order $n$ and degree $m$ at each grid point $(\theta_q, \phi_q)$:

\begin{equation}
\mathbf{y}_{nm} = [Y_{nm}(\theta_1, \phi_1), \dots, Y_{nm}(\theta_Q, \phi_Q)]^T
\label{eq:spherical_harmonics_vector}
\end{equation}

\noindent We use the real SN3D spherical harmonics basis described in \citep{gorzel2019efficient}. The beamforming weights are independent of the input signal. They can be computed offline and applied across STFT frames.

The beamforming weights are computed for each array configuration. They are applied to each STFT frame as described above. The inverse-STFT is applied to compute the time-domain estimated FOA signal before feature extraction and encoding.

\section{Beamforming Artifacts and Array Selection}
\label{appendix:beamforming_artifacts}

We have observed that ASM beamforming often introduces spectral coloration to the encoded signal, which frequently manifests as a low-pass filtering effect. This high-frequency attenuation arises from two compounding factors. First, as the acoustic wavelength becomes comparable to or smaller than the physical distance between the array's microphones, spatial aliasing prevents the accurate reconstruction of high-frequency spatial cues \citep{rafaely2015fundamentals}. Second, to manage numerical instability, we apply Tikhonov regularization during the computation of the beamforming weights, which can penalize the output power at high frequencies.

\begin{figure}[t]
    \centering
    \includegraphics[width=\linewidth]{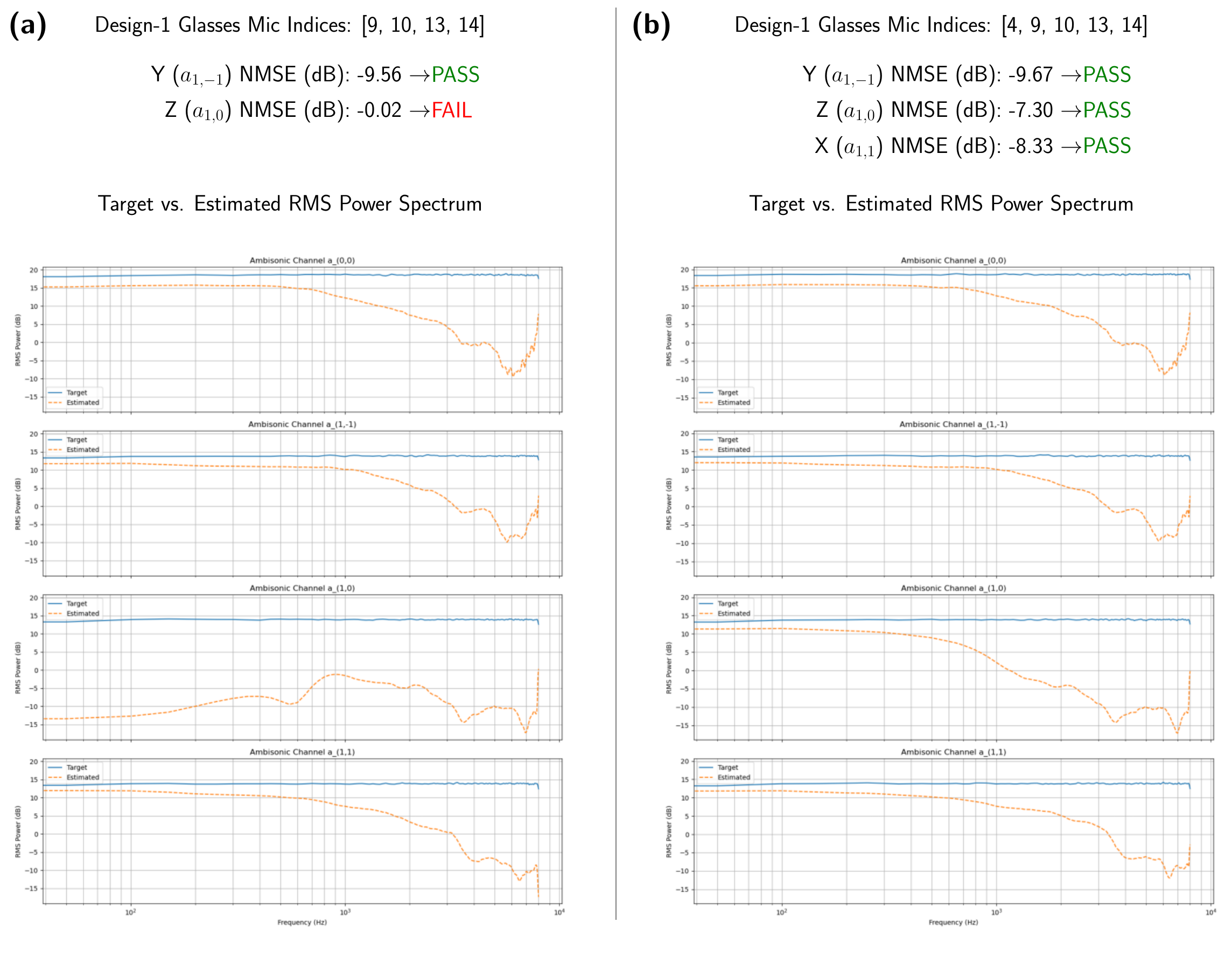}
    \caption{\small\textbf{Ambisonic Signal Matching Simulation Results.} (a) A 4-microphone array configuration on the Design 1 smart glasses (the microphone indices are visualized in Figure \ref{fig:Design-1_array}), with microphones located at the front and rear of both the left and right temples. Because these microphones are strictly coplanar, the elevation (Z) channel cannot be accurately reconstructed, resulting in significant attenuation across the RMS power spectrum and failure to meet our Normalized Mean Squared Error (NMSE) selection criteria. (b) A 5-microphone configuration that adds a single microphone on the bottom-right rim of the glasses. This addition introduces necessary Z-axis displacement, allowing the array to reconstruct the elevation channel sufficiently and pass the selection criteria. In both configurations, the inherent low-pass filtering effect of ASM beamforming is evident across most reconstructed channels.}
    \label{fig:asm_simulation_figure}
\end{figure}

Furthermore, certain FOA channels may suffer from significant attenuation and poor reconstruction if the microphones lack sufficient displacement along the corresponding axis. For instance, if the microphones are predominantly coplanar on the horizontal plane, the elevation (Z) channel cannot be accurately reconstructed. Both the low-pass filtering and the channel attenuation artifacts are illustrated in Figure \ref{fig:asm_simulation_figure}.

There are numerous ways to combine 4 to 9 microphones across both smart glasses designs. However, not all microphone configurations are conducive to estimating the FOA signal (e.g., arrangements that are strictly coplanar). We leverage our simulation study to systematically select configurations capable of adequately reconstructing the FOA signal. This selection process is essential to set the encoder up for success; if we included arrays incapable of capturing spatial cues, the model would be forced to learn from ambiguous examples lacking directional information, rendering the task practically impossible and destabilizing the training process.

We conduct a search over many 4- to 9-microphone configurations from both glasses designs. For each configuration, we execute the simulation study described in Appendix \ref{appendix:asm_simulation} and select arrays that meet the following criteria: (i) \textbf{Azimuth Channels (X and Y):} NMSE $\le -5$ dB below 500 Hz. (ii) \textbf{Elevation Channel (Z):} NMSE $\le -1$ dB below 100 Hz. Through this search, we identified 1211 arrays that successfully passed these criteria.

These thresholds were empirically determined to select array configurations that robustly preserve azimuth cues while maintaining adequate elevation cues. The constraints on the Z-channel are relaxed relative to the X and Y channels due to the physical form factor of the glasses, where microphones are predominantly coplanar. Specifically, there is more spatial displacement between microphones along the X and Y axes than along the Z axis, making the reconstruction of elevation cues inherently more challenging. Figure \ref{fig:asm_simulation_figure} contrasts a passing array configuration against a failing configuration hindered by coplanar microphone placement.

\section{\samethod{} Training and Evaluation}
\label{appendix:foacoder_evaluation}

\subsection{Training Objectives}
\label{appendix:foacoder_pretraining_objectives}
The spatial audio encoder is pre-trained using a multi-task learning paradigm that forces the representations to encapsulate both the temporal presence of sound and its precise geometric location in 3D space. We attach two parallel Multi-Layer Perceptron (MLP) heads to the backbone sequence $\mathbf{H}$: a Voice Activity Detection (VAD) head and a Sound Event Localization and Detection (SELD) head.

\noindent\textbf{1. Voice Activity Detection (VAD) Objective.}
The VAD head projects the hidden representations to a single logit per time step to predict speech presence. The output $\hat{\mathbf{y}}^{\text{vad}} \in \mathbb{R}^{B \times L}$ represents the unnormalized logits. We optimize this using a standard Binary Cross-Entropy (BCE) loss across the sequence length $L$. Let $m_t$ be 1 if a frame is valid (not padding) and 0 if it is padding. The VAD loss is computed as:
\begin{equation}
\mathcal{L}_{\text{VAD}} = - \frac{1}{L_{\text{not-padded}}} \sum_{t=1}^{L} m_t [ y_{t}^{\text{vad}} \log(\sigma(\hat{y}_{t}^{\text{vad}})) + (1 - y_{t}^{\text{vad}}) \log(1 - \sigma(\hat{y}_{t}^{\text{vad}}))]
\end{equation}
where $\sigma(\cdot)$ is the sigmoid function and $y_{t}^{\text{vad}} \in \{0, 1\}$ is the ground-truth VAD label for frame $t$, which specifically corresponds to self/wearer speech.

\noindent\textbf{2. Sound Event Localization and Detection (SELD) Objective.}
To handle multiple overlapping sound sources of the same class, we format the model's output targets using the Multi-Activity Coupled Cartesian Direction of Arrival (Multi-ACCDOA) format \citep{shimada2022multi}. Furthermore, to address the track permutation problem inherent in multi-track outputs, we train the network using the Location-based Training (LBT) paradigm \citep{scheibler2022sorting}. Under this framework, to learn precise geometric coordinates, the SELD head projects the hidden representations into a multi-class, multi-track Cartesian coordinate space. For $N_c=2$ distinct sound classes and a maximum of $N_t=3$ overlapping tracks per class, the SELD head outputs $\hat{\mathbf{P}} \in \mathbb{R}^{B \times L \times N_c \times N_t \times 3}$, directly regressing the 3D Cartesian coordinates $(x, y, z)$ per frame.

For a given frame and class index, if there are fewer sources than the number of tracks, we perform repeat padding. Specifically, the method fills any remaining unoccupied tracks by duplicating the last present event for that class and frame. If there are no sources that belong to that class, all tracks are simply filled with zeros. When multiple sources of the same class are present, they are sorted amongst the available tracks based on their Cartesian distance to the origin, ensuring consistent target assignment.

Crucially, the $(x, y, z)$ coordinates we regress on are all unit vectors. When no sources are present for a class, $(x, y, z) = (0, 0, 0)$; otherwise the norm $\sqrt{x^2 + y^2 + z^2} = 1$, making the norm function as an implicit source detection confidence score. We regress on direction only, omitting explicit distance prediction.

We formulate the localization objective using a standard Mean Squared Error (MSE) loss. Let $m_t$ be the mask (1 if the frame is valid, 0 if it is padding). Padded frames are not included in the loss. The SELD loss is computed as:
\begin{equation}
\mathcal{L}_{\text{SELD}} = \frac{1}{K_{\text{act.}}} \sum_{t=1}^{L} \sum_{c=1}^{N_c} \sum_{k=1}^{N_t} m_t \left\| \mathbf{p}_{t,c,k} - \hat{\mathbf{p}}_{t,c,k} \right\|_2^2
\end{equation}
where $\mathbf{p}_{t,c,k}$ and $\hat{\mathbf{p}}_{t,c,k}$ are the ground-truth and predicted 3D Cartesian coordinates, respectively, and $K_{\text{act.}} = \sum_{t,c,k} m_t$ is the total number of valid frames to normalize the loss.

\noindent\textbf{Joint Objective.}
The total pre-training loss for the spatial audio encoder is the weighted sum of the VAD and SELD objectives:
\begin{equation}
\mathcal{L}_{\text{encoder}} = \lambda_{\text{VAD}} \mathcal{L}_{\text{VAD}} + \lambda_{\text{SELD}} \mathcal{L}_{\text{SELD}}
\end{equation}
where $\lambda_{\text{VAD}} = 1.0$ and $\lambda_{\text{SELD}} = 1.0$ are hyperparameters controlling the relative contribution of each task. Upon completion of this pre-training phase, the task-specific heads are discarded, the rest of the architecture is frozen and the embedding sequence $\mathbf{H}$ is routed into the spatial projector of the main LALM architecture.

\subsection{Training Setup for \samethod.}
\label{appendix:foacoder_training_setup}
We train \samethod{} for 2,000,000 steps with a batch size of 128 on 8$\times$NVIDIA H100 (80GB) GPUs. We employ an AdamW optimizer configured with $\beta_1=0.9$, $\beta_2=0.999$, and a weight decay of $1\times10^{-2}$. Each optimizer uses a warmup cosine decay learning rate schedule with an initial learning rate of $1\times10^{-6}$, warming up over 500,000 steps to a peak learning rate of $2\times10^{-5}$, and decaying over 2,000,000 steps to a final learning rate of $1\times10^{-6}$.

\subsection{Evaluation Protocol}
\label{sec:foacoder_eval}
We evaluate the encoder on the evaluation set of our synthetic spatial audio dataset, which consists of 156,000 spatial audio examples. For each example, we have metadata such as the start and end time of each sound source, the type and description of each sound (such as speech or dog barking), and the spatial location for external sound sources. This metadata is used to derive ground truth for our evaluations. During evaluation, the encoder strictly predicts two sound classes: speech and non-speech.

\noindent\textbf{Evaluation Metrics.} Our encoder evaluation metrics fall under two categories: Sound Event Localization and Detection (SELD) metrics and Wearer Voice Activity Detection (VAD) metrics.

\textit{SELD Metrics:} We evaluate detection and localization capabilities using standard DCASE metrics \citep{mesaros2019joint, politis2020overview}: Angular Error (the average angular distance between correctly paired reference and predicted points), Error Rate (a location-dependent error ratio of substitutions, deletions, and insertions over total references), F-score (detection F-score constrained by a strict location-dependent angular threshold), Localization Recall (true-positive recall of target events correctly localized), and an aggregated SELD Score (a macro score summarizing performance). To compute these errors, we establish a correspondence between the predicted and ground-truth sources frame-by-frame by assigning target matches that minimize angular distances using the Hungarian algorithm. The metric computations were adapted from the DCASE 2024 SELD Evaluator Baseline repository \citep{aparicio2024baseline}\footnote{\url{https://github.com/partha2409/DCASE2024_seld_baseline}}.

\textit{Wearer VAD Metrics:} VAD in this context specifically refers to Wearer VAD (detecting the voice activity of the user wearing the device). We report the aggregate mean of several standard binary metrics computed on independent acoustic timeframes: Accuracy, Precision, Recall, and F1-score (evaluating timeframes where the unactivated logits cross the activation threshold), as well as the AUC measuring separation power acting directly on probabilities. Any padded portion of the 30-second audio clip is excluded from the calculation. Furthermore, these detection metrics are computed per example and then averaged across examples.

\noindent\textbf{1s Context \samethod{}.} In addition to the baseline 30-second context window encoder, we trained a 1-second context window version, keeping the rest of the architecture and its number of parameters similar. While dynamic acoustic scenes are not within the scope of this paper, a smaller context-window model can be more easily extended to dynamic environments in future work. With a shorter context window, the model can generate embeddings for brief periods where the acoustic scene is functionally static. To run inference on the full 30-second spatial audio clips, we apply a sliding window inference approach over the audio representation. We did not utilize this 1-second context encoder in any further downstream experiments.

\noindent\textbf{SELDNet Baseline.} We also evaluate SELDNet, the DCASE 2024 baseline method \citep{aparicio2024baseline}, for comparison. We resample our 16 kHz spatial audio dataset to 24 kHz, which is the native sampling rate accepted by SELDNet. SELDNet initially outputs 13 sound classes; we consolidate these into our binary taxonomy by mapping classes 0 and 1 (female and male speech) into a unified ``speech'' class, and grouping all remaining classes into ``non-speech.'' Because SELDNet does not natively estimate Wearer VAD, we heuristically derive it from the model outputs. Specifically, we threshold the spatial distance of any detected speech at 0.15~m to determine if the speech physically originates close enough to be the wearer.

\subsection{Results}
\label{sec:foacoder_results}

\begin{table*}[t]
\centering
\caption{Encoder evaluation results. Best values in each column are indicated in \textbf{bold}.}
\label{tab:encoder_eval}
\resizebox{\linewidth}{!}{
\begin{tabular}{llrrrrr}
\toprule
\multicolumn{7}{c}{\textbf{(a) DCASE Metrics}} \\
\midrule
\textbf{Model} & \textbf{Parameter Count} & \textbf{Angular Error ($^\circ$) $\downarrow$} & \textbf{Error Rate $\downarrow$} & \textbf{F-score $\uparrow$} & \textbf{Localization Recall $\uparrow$} & \textbf{SELD Score $\downarrow$} \\
\midrule
SELDNet & 0.74M & 44.01 & 1.20 & 12.62 & 32.29 & 0.75 \\
\samethod{} 1s context & 46.03M & 15.84 & 0.41 & 63.43 & 75.87 & 0.28 \\
\samethod{} 30s context & 46.18M & \textbf{15.06} & \textbf{0.34} & \textbf{70.46} & \textbf{86.50} & \textbf{0.21} \\
\bottomrule
\end{tabular}
}

\vspace{1em}

\resizebox{\linewidth}{!}{
\begin{tabular}{llrrrrr}
\toprule
\multicolumn{7}{c}{\textbf{(b) VAD Metrics}} \\
\midrule
\textbf{Model} & \textbf{Parameter Count} & \textbf{Accuracy $\uparrow$} & \textbf{AUC $\uparrow$} & \textbf{F1 $\uparrow$} & \textbf{Precision $\uparrow$} & \textbf{Recall $\uparrow$} \\
\midrule
SELDNet & 0.74M & 75.44 & 50.02 & 0.07 & 67.39 & 0.04 \\
\samethod{} 1s context & 46.03M & \textbf{99.84} & \textbf{99.96} & 98.12 & 98.32 & 99.37 \\
\samethod{} 30s context & 46.18M & 99.82 & 99.95 & \textbf{98.53} & \textbf{98.58} & \textbf{99.41} \\
\bottomrule
\end{tabular}
}
\end{table*}

The results of the encoder evaluation are shown in Table~\ref{tab:encoder_eval}. Overall, the \samethod{} models perform significantly better than the SELDNet baseline across both DCASE and VAD metrics. We hypothesize that one possible reason for this performance gap in spatial localization (DCASE metrics) is the difference in microphone array geometries used to create the training data. Notably, we did not fine-tune the SELDNet baseline on our synthetic dataset. It was trained entirely on the STARSS23 dataset, a real dataset recorded using a highly favorable 32-channel spherical microphone array converted to FOA \citep{shimada2023starss23, aparicio2024baseline}, which produces high-quality signals. In contrast, the \samethod{} was trained on our synthetic dataset with signals beamformed from irregular arrays derived from glasses geometries. This irregularity introduces non-idealities in the resultant ambisonic signals, preparing the model for the non-ideal signals encountered at inference. A second likely contributing factor is that the \samethod{} models possess significantly more parameters ($\sim$46.18M and $\sim$46.03M) compared to the SELDNet baseline ($\sim$0.74M).

While the 30-second context window encoder displays moderately better performance and was ultimately selected for the \method{} spatial context, the 1-second encoder demonstrates promising performance that might be investigated in future work for addressing dynamic acoustic scenes. Furthermore, despite having a similar parameter footprint to the 30-second encoder, the 1-second encoder drastically reduces inference compute and latency. Because these models are based on the Transformer architecture, which scales quadratically with sequence length, processing sequences that are 30 times shorter yields significant computational savings.

We note that the VAD metrics achieved by the \samethod{} models are remarkably high, which is likely because wearer Voice Activity Detection is a relatively easier problem compared to Sound Event Localization and Detection (SELD). \samethod{} also significantly outperforms the SELDNet baseline on these VAD metrics. In addition to the training data and capacity advantages mentioned above, another reason for the performance gap might be that \samethod{} models are explicitly trained with a dedicated wearer Voice Activity Detection (VAD) objective, whereas the baseline was not tailored to natively estimate this attribute. Because SELDNet was not fine-tuned on our dataset and possesses a significantly smaller parameter footprint, this comparison is not intended as a direct architectural benchmark. Rather, we utilize the SELDNet baseline primarily as a diagnostic reference to demonstrate the substantial acoustic complexity of our synthetic spatial audio scenes.

\section{Glasses Microphone Arrays}
\label{appendix:glasses_arrays}
\begin{figure}[t]
    \centering
    \begin{subfigure}[t]{0.485\textwidth}
        \centering
        \includegraphics[width=\linewidth]{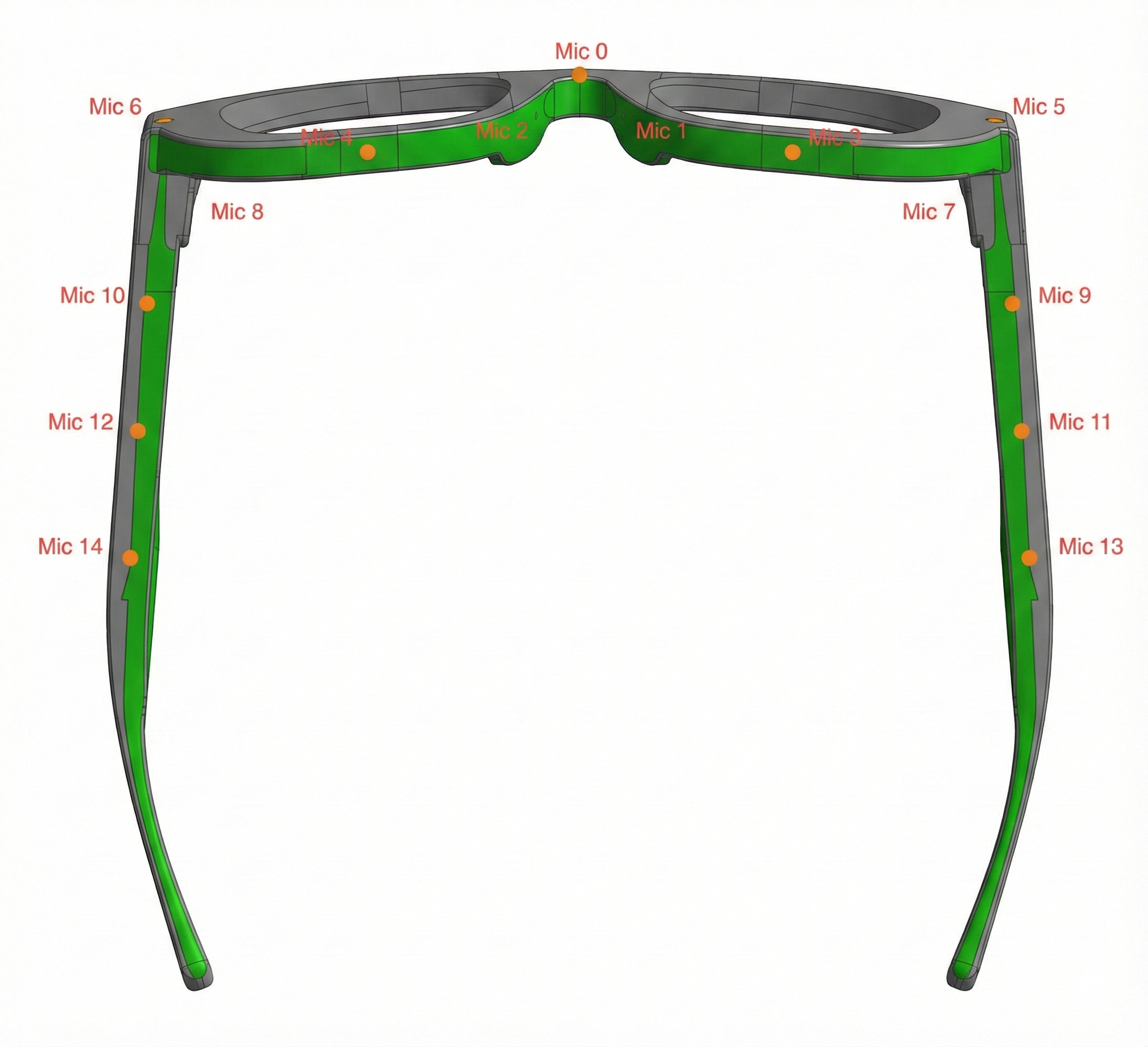}
        \caption{Design 1 microphone topology.}
        \label{fig:Design-1_array}
    \end{subfigure}
    \hfill
    \begin{subfigure}[t]{0.485\textwidth}
        \centering
        \includegraphics[width=\linewidth]{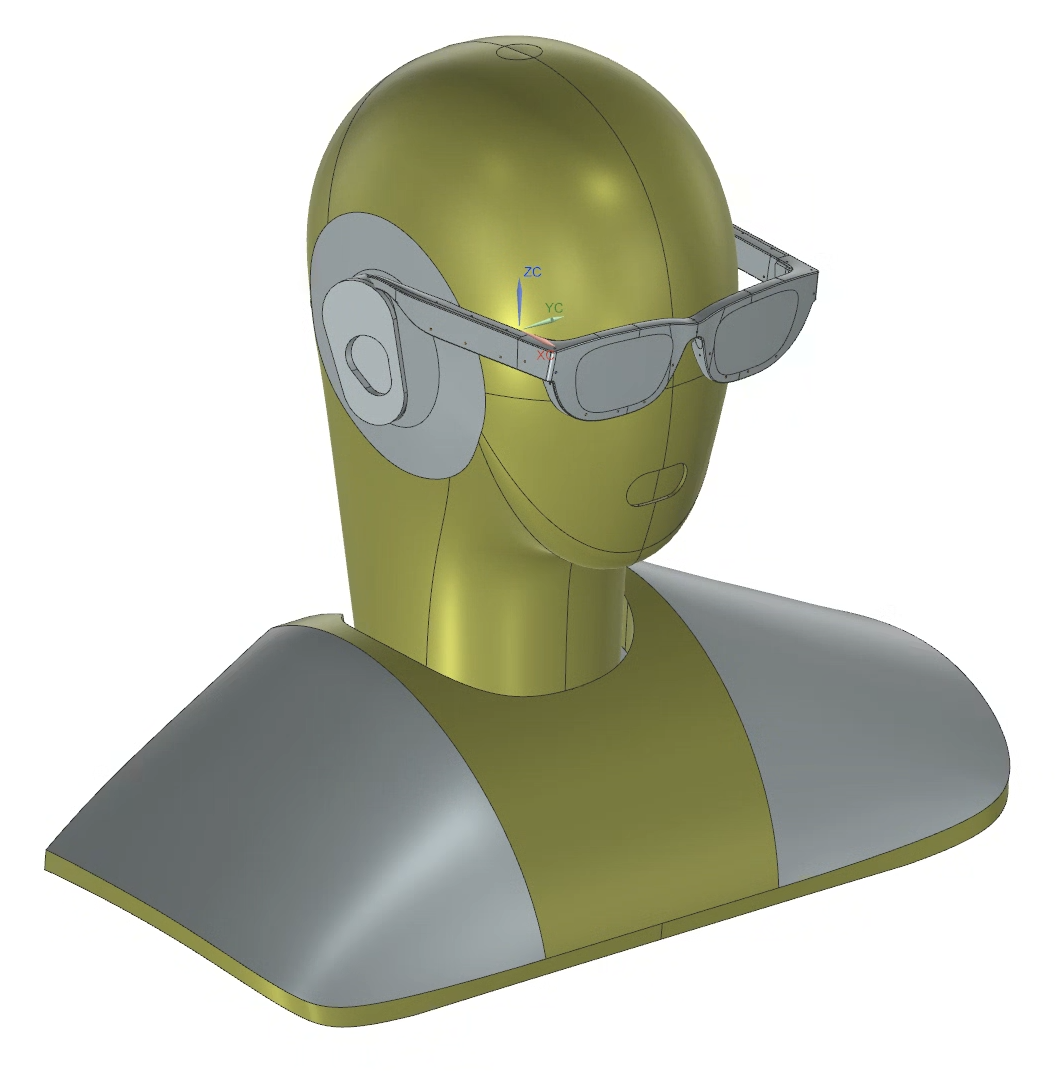}
        \caption{Design 1 glasses fitted on the HAT model.}
        \label{fig:hats_with_glasses}
    \end{subfigure}

    \vspace{0.75em}

    \begin{subfigure}[t]{0.485\textwidth}
        \centering
        \includegraphics[width=\linewidth]{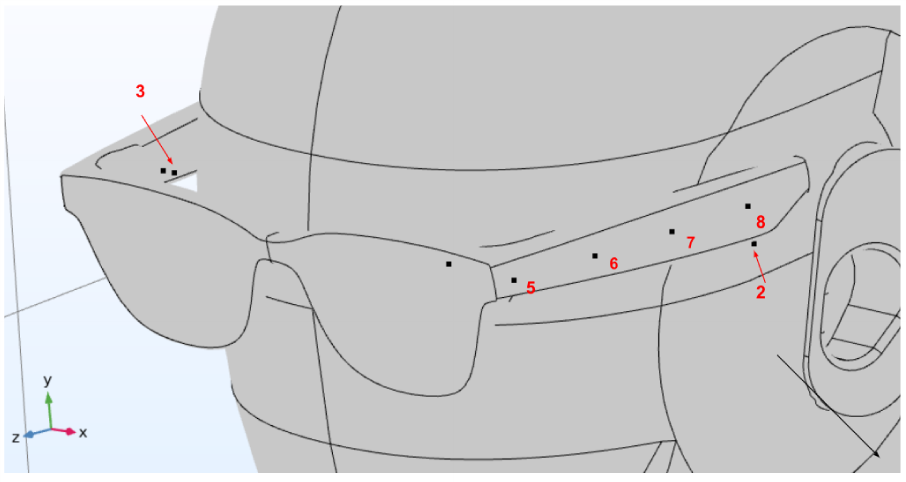}
        \caption{Design 2 microphone topology, left view.}
        \label{fig:design_2_array}
        \label{fig:design_2_array_1}
    \end{subfigure}
    \hfill
    \begin{subfigure}[t]{0.485\textwidth}
        \centering
        \includegraphics[width=\linewidth]{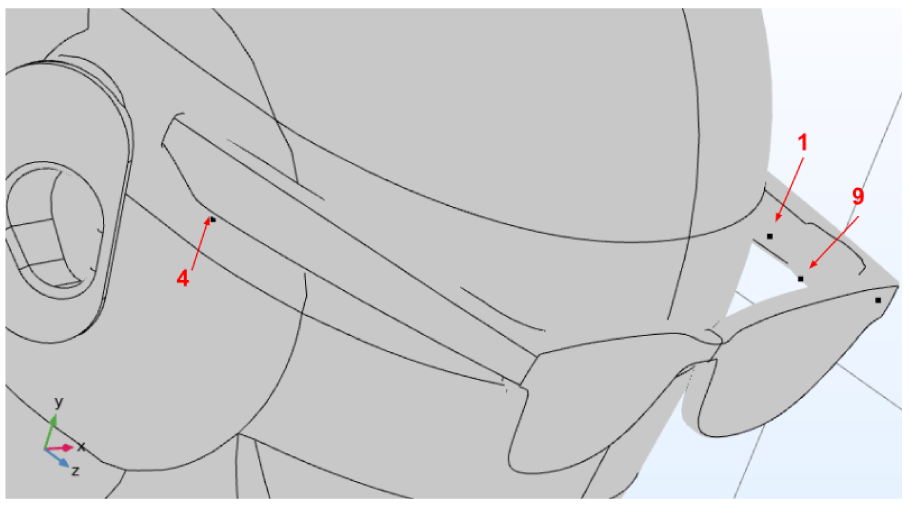}
        \caption{Design 2 microphone topology, right view.}
        \label{fig:design_2_array_2}
    \end{subfigure}
    \caption{\small\textbf{Smart-glasses microphone array designs and simulation geometry.} (a) Microphone placement on the Design 1 glasses array. (b) The Design 1 glasses fitted on a HEAD acoustics HMS II head-and-torso (HAT) model for array transfer function simulation. (c--d) Complementary views of the Design 2 glasses array topology. In both topology designs, numbered red arrows identify the microphone positions referenced in the array-selection analysis.}
    \label{fig:design_1_array_and_hats}
    \vspace{-1em}
\end{figure}

As mentioned in Subsection~\ref{subsubsec:array_selection}, the topology of glasses microphone array is shown in Fig.~\ref{fig:Design-1_array}, Fig.~\ref{fig:design_2_array}.

\section{Array Transfer Function Simulation Details}
\label{appendix:atf_simulation_details}

Let $H_{ff,i}(\theta, \phi, f) \in \mathbb{C}$ denote the far-field-to-microphone transfer function for microphone $i$, corresponding to a source at elevation angle $\theta$ and azimuth angle $\phi$ at frequency $f$. This function characterizes how sound emitted from a specific directional source is received by microphone $i$. Similarly, the mouth-to-microphone transfer function, $H_{m,i}(f) \in \mathbb{C}$, characterizes how speech emitted from the wearer's mouth is received by microphone $i$. We aim to run acoustic simulations to obtain these transfer functions given the CAD model of a device.

\subsection{Simulation Setup}
The 3D model of the glasses is fitted onto a HEAD acoustics HMS II head and torso (HAT) model, which does not include a nose, as shown in Figure \ref{fig:hats_with_glasses}. This combined geometry is imported into COMSOL, with the HAT positioned such that the midpoint of the interaural axis lies at the origin.

Simulating the far-field-to-microphone pressure responses for numerous external sources individually is computationally expensive. Therefore, we exploit the acoustic reciprocity principle \citep{kinsler2000fundamentals} to swap the positions of the sources and receivers. Instead of placing sound sources on a sphere with a 1 m radius, a single sound source is placed directly on the microphone surface. We use the Boundary Element Method (BEM) to simulate the resulting pressure response across a surrounding sphere with a 1 m radius. We export these pressure responses on a spatial grid of 12,602 points on the sphere, ranging from $0^\circ$ to $359^\circ$ in azimuth ($1^\circ$ increments) and $-90^\circ$ to $90^\circ$ in elevation ($5^\circ$ increments). We simulate 128 linear frequency bins from 62.5 Hz to 8000 Hz. These computed pressure responses are mathematically equivalent to simulating each far-field source individually.

Similarly, for the mouth-to-microphone pressure responses, we place a sound source directly on the microphone surface. Using BEM, we calculate the resulting sound pressure across the surface of the mouth. This sound pressure is then spatially averaged over the mouth surface to obtain the final mouth-to-microphone pressure response. We use the same frequency grid as described above for these simulations.

\subsection{Post-processing}

The simulated far-field-to-microphone and mouth-to-microphone transfer functions are referenced against analytically calculated free-field pressure responses, following \citep{brinkmann2023recent}. A DC value ($0\text{ Hz}$) is added to each transfer function by duplicating the value from the lowest simulated frequency bin. 

Far-field transfer functions are normalized by scaling all complex responses relative to the maximum magnitude recorded across all directions and microphones at DC. Similarly, mouth transfer functions are normalized so that the microphone closest to the mouth has a response magnitude of 0 dB at DC. These schemes are designed to ensure consistency between different sets of simulated array transfer functions.

At the Nyquist frequency ($f_{\text{sampling}} = 16\text{ kHz}$, $f_{\text{Nyquist}} = f_{\text{sampling}}/2 = 8\text{ kHz}$), the transfer function must be real-valued; otherwise, the inverse-FFT will yield a complex time-domain signal. Following \citep{brinkmann2023recent}, we enforce this by replacing the complex pressure value at the Nyquist bin with its absolute magnitude. The transfer functions are then converted into time-domain impulse responses using the inverse-FFT. To avoid ringing artifacts and to ensure the impulse response decays to zero, a Hann window fade-out is applied to the end of each impulse response. We save these impulse responses to SOFA files \citep{sofa_general} for convenient loading into our room impulse response generation pipeline.

\section{Ambisonic Signal Matching Simulation Study}
\label{appendix:asm_simulation}

We use the simulation study for two primary purposes: first, to demonstrate the acoustic artifacts induced by Ambisonic Signal Matching (ASM) beamforming, and second, to establish our array selection methodology. By computing an objective performance metric through these simulations, we systematically eliminate microphone arrays that fail to sufficiently reconstruct the First-Order Ambisonics (FOA) audio.

Our simulation study closely follows the methodology of \citep{gayer2024ambisonics}. The signal model we use for the simulation is:
\begin{equation}
\mathbf{x}(f) = \mathbf{V}(f)\mathbf{s}(f)
\label{eq:asm_simulation_signal_model}
\end{equation}

\noindent where $\mathbf{x}(f) \in \mathbb{C}^{N_{\text{mics}}}$ is the received multichannel signal and $\mathbf{V}(f)$ is the steering matrix, as defined in Appendix \ref{appendix:asm_beamforming_details}. The vector $\mathbf{s}(f) = [s_1(f), \dots, s_Q(f)]^T \in \mathbb{C}^Q$ contains the source signals emitted by $Q$ far-field sources.  We use $Q=240$ far-field sources that form a t-design spatial grid so that the sources are distributed uniformly on the unit sphere. We omit the microphone noise term from this model for simplicity.

The full target Ambisonics signal vector due to $\mathbf{s}(f)$ is computed as:

\begin{equation}
\mathbf{a}(f) = \mathbf{Y}_{\text{sh}}^H \mathbf{s}(f)
\label{eq:target_ambisonic_signal}
\end{equation}

\noindent Here, $\mathbf{a}(f) \in \mathbb{C}^{(N_a + 1)^2}$ is a vector containing all $(N_a + 1)^2$ Ambisonic channels, and $\mathbf{Y}_{\text{sh}} = [\mathbf{y}_{00}, \mathbf{y}_{1-1}, \mathbf{y}_{10}, \mathbf{y}_{11}] \in \mathbb{R}^{Q \times (N_a + 1)^2}$ is the spherical harmonics matrix, where each column is defined according to Equation \ref{eq:spherical_harmonics_vector}. Let $a_{nm}(f)$ denote the individual scalar component corresponding to Ambisonic order $n$ and degree $m$.

We simulate multiple frames (20 seconds total duration) of the signal $\mathbf{s}(f)$ directly in the STFT domain. Both the real and imaginary components of each source signal $s_i(f)$ are drawn from a standard normal distribution $\mathcal{N}(0, 1)$. To ensure that the complex variance of each frequency bin is exactly 1, the sum of the real and imaginary parts is scaled by a factor of $1/\sqrt{2}$. These signals are generated independently for every source direction, time frame, and frequency bin. Next, for each frame, using the steering matrix $\mathbf{V}(f)$, we compute the simulated received multichannel signal $\mathbf{x}(f)$ via Equation \ref{eq:asm_simulation_signal_model}. We compute the beamforming weights using Equation \ref{eq:beamformer_weights} and apply these weights to each frame of the input signal using Equation \ref{eq:beamformer_apply} to obtain the estimated FOA scalar component $\hat{a}_{nm}(f)$. 

To evaluate performance, we compute the normalized mean squared error (NMSE) for each frequency bin $f$ and specific Ambisonic component ($n, m$):

\begin{equation}
\text{NMSE}_{nm}(f) = \frac{\mathbb{E} \left[ \left| \hat{a}_{nm}(f) - a_{nm}(f) \right|^2 \right]}{\mathbb{E} \left[ \left| a_{nm}(f) \right|^2 \right]}
\end{equation}

\noindent where the expectation operator $\mathbb{E}[\cdot]$ denotes averaging over the simulated time frames.

Furthermore, we compute and plot the Root Mean Square (RMS) power spectrum (in dB) of the target signal for the Ambisonic component of order $n$ and degree $m$:

\begin{equation*}
\text{RMS}_{nm}^{\text{dB}}(f) = 10 \log_{10} \left( \mathbb{E} \left[ \left| a_{nm}(f) \right|^2 \right] \right)
\end{equation*}

\noindent We identically compute and plot this metric for the estimated signal $\hat{a}_{nm}(f)$. The RMS power spectrum is evaluated to observe the spectral coloration and artifacts introduced to the audio signal by the beamforming process. The NMSE metric serves as a foundational component of our array selection criteria. Results from the simulation and the array selection methodology are presented in Appendix \ref{appendix:beamforming_artifacts}.

\section{Room Impulse Response Simulation Details}
\label{appendix:rir_simulation_details}

\subsection{Simulating External Sources}
We conduct the simulations in the frequency domain. Let $P$ denote the number of image source paths from a given sound source to the device. We compute the transfer function of the received signal at microphone $i$ and frequency $f$ as:

\begin{equation}
r_{\text{ex},i}(f) = \sum_{p=1}^{P} I_p(f)e^{-j2\pi f \left(\tau_p - \frac{d_{ff}}{c}\right)}H_{ff,i}(\theta_p, \phi_p, f)
\label{eq:rir_simulation_external_source}
\end{equation}

\noindent where, for each path $p$, $I_p(f) \in \mathbb{C}$ is a multiplier governing distance attenuation and wall filtering, $\tau_p$ is the travel delay, and $(\theta_p, \phi_p)$ is the direction of arrival. Here, $d_{ff} = 1\text{ m}$ is the radius of the sphere where the far-field sources were placed during the ATF simulations, and $c = 343\text{ m/s}$ is the speed of sound. The term $-\frac{d_{ff}}{c}$ applies a negative time shift to undo the propagation delay inherent in the ATF simulations, which is induced by the distance from the far-field sources to the microphone array.

During the simulations, we do not have access to the exact continuous transfer function $H_{ff,i}(\theta_p, \phi_p, f)$ for arbitrary directions $(\theta_p, \phi_p)$. Instead, we approximate it by selecting the closest available direction from our pre-computed spatial grid.

\subsection{Simulating the Self Speech Source}

The room impulse response at microphone $i$, generated by the wearer's mouth, is modeled as follows (the microphone index $i$ is omitted for brevity):

\begin{equation}
r_{m}(t) = h_{m}(t) + h_{\text{reflections}}(t)
\label{eq:mouth_rir}
\end{equation}

\noindent where $h_{m}(t) \in \mathbb{R}$ is the simulated mouth-to-microphone impulse response (the accurate direct path), and $h_{\text{reflections}}(t) \in \mathbb{R}$ is the impulse response containing exclusively the room reflections of the wearer's speech. 

To compute $h_{\text{reflections}}(t) \in \mathbb{R}$, a point source is placed at the mouth position, and RIRs are simulated using Equation \ref{eq:rir_simulation_external_source}, treating the mouth as an external far-field sound source. This initial RIR captures both the room reflections and an inaccurate direct path, since the mouth is erroneously treated as a far-field source:

\begin{align*}
r_{\text{point-source-mouth}}(t) &= h_{\text{inaccurate-direct-path}}(t) \\
&\quad + h_{\text{reflections}}(t)
\end{align*}

\noindent To isolate and remove this error, we repeat the simulation using the exact same point source, but within an entirely anechoic room (completely non-reflective walls). This yields solely the inaccurate direct path, $h_{\text{inaccurate-direct-path}}(t)$. Subtracting this anechoic response from the full room simulation isolates the pure reflections:

\begin{align*}
h_{\text{reflections}}(t) &= r_{\text{point-source-mouth}}(t) \\
&\quad - h_{\text{inaccurate-direct-path}}(t)
\end{align*}

\noindent The final RIR for the mouth source is then constructed using Equation \ref{eq:mouth_rir}.

\subsection{Simulation Parameters}

The following parameters are randomized during the room generation process:
\begin{itemize}
    \item \textbf{Room Dimensions}: The width ($X$) and length ($Y$) are independently and uniformly sampled between $2.0$ and $15.0$ meters, and the height ($Z$) between $3.0$ and $6.0$ meters.
    \item \textbf{Device Wearer Position}: The device wearer is placed randomly within the room.
    \item \textbf{External Sound Source Distance}: The external sound sources are placed at a radial distance from the device wearer, sampled from a uniform distribution between $0.6$ and $12.0$ meters.
    \item \textbf{Array Type}: The system uniformly samples one of the 1211 possible array configurations.
    \item \textbf{Wall Material}: Acoustic materials for the floor and ceiling are sampled independently from predefined categorical lists. All four vertical side walls are assigned the same randomly selected wall material.
    \item \textbf{Wall Filter Gain}: A global scalar gain is uniformly sampled between $0.75$ and $0.95$ and applied to all wall reflection filters to adjust reflectivity.
\end{itemize}

To ensure realistic geometries and prevent simulation artifacts and errors, we enforce several constraints on the sampled room:

\begin{itemize}
    \item \textbf{Device Wearer to Walls}: The wearer must be inside the room and at least $0.2$ meters from any wall.
    \item \textbf{External Sources to Walls}: External sound sources must be inside the room and at least $0.1$ meters from any wall.
\end{itemize}

\section{Mixing Parameters}
\label{appendix:mixing_parameters}

Each spatial audio example features up to three distinct sources. Source 0 represents the wearer's self-speech, while Sources 1 and 2 represent external sound events or bystander speech. Because each source is activated independently with a probability of $0.5$, approximately $12.5\%$ of the dataset consists of complete silence, which intentionally serves as negative examples for the model. The randomization scheme for the mixing parameters of each source is detailed below.

\begin{itemize}
    \item \textbf{Source 0 (Self-speech):}
    \begin{itemize}
        \item \textbf{Presence Probability:} $P(\text{presence}) = 0.5$.
        \item \textbf{Dataset Selection:} Always sampled from \textit{LibriTTS}.
        \item \textbf{Gain Distribution:} Uniformly distributed as $\mathcal{U}(-9.0, 0.0)$ dB.
        \item \textbf{Delay Distribution:} Uniformly distributed as $\mathcal{U}(0.0, 10.0)$ seconds.
    \end{itemize}

    \item \textbf{Source 1 (External Source):}
    \begin{itemize}
        \item \textbf{Presence Probability:} $P(\text{presence}) = 0.5$.
        \item \textbf{Dataset Selection:} Sampled from \textit{LibriTTS} ($p=0.8$), \textit{AudioSet} ($p=0.1$), or \textit{Freesound} ($p=0.1$).
        \item \textbf{Gain Distribution:} Uniformly distributed as $\mathcal{U}(-6.0, 3.0)$ dB.
        \item \textbf{Delay Distribution:} Uniformly distributed as $\mathcal{U}(0.0, 10.0)$ seconds.
    \end{itemize}

    \item \textbf{Source 2 (External Source):}
    \begin{itemize}
        \item \textbf{Presence Probability:} $P(\text{presence}) = 0.5$.
        \item \textbf{Dataset Selection:} Sampled from \textit{AudioSet} ($p=0.6$), \textit{LibriTTS} ($p=0.2$), or \textit{Freesound} ($p=0.2$).
        \item \textbf{Gain Distribution:} Uniformly distributed as $\mathcal{U}(-6.0, 3.0)$ dB.
        \item \textbf{Delay Distribution:} Uniformly distributed as $\mathcal{U}(0.0, 10.0)$ seconds.
    \end{itemize}
\end{itemize}

\section{Question and Answer Generation Details and Examples}
\label{sec:qa_examples}

\noindent\textbf{Generation details.}
For any given example or scene, the valid tasks are determined based on what the scene includes. Specifically, task validity depends on the existence of wearer speech, external speech, and external non-speech sources. Table~\ref{tab:task_validity_matrix} illustrates this task validity matrix. Once the valid tasks are determined based on this matrix, we sample tasks uniformly for each example without replacement from this valid set. For each sampled task, the prompt is paraphrased from a set collection of prompts. For tasks that include source selection (Selective Sound Localization (Task 2) and Targeted External Speech Transcription (Task 6)), the target sound description or direction is chosen randomly based on the available sound descriptions and directions in the scene. Importantly, these tasks may incorporate negative queries by querying an absent sound description or direction, producing \texttt{\{"sources":[]\}}.

\begin{table*}[htb]
\centering
\scriptsize
\caption{Task validity matrix based on scene content. A checkmark (\cmark) indicates that the presence of the specified sound source satisfies the requirements for the corresponding task to be valid.}
\label{tab:task_validity_matrix}
\setlength{\tabcolsep}{4pt}
\resizebox{\linewidth}{!}{
\begin{tabular}{@{}lcccccc@{}}
\toprule
\textbf{Scene Content} & 
\begin{tabular}[c]{@{}c@{}}\textbf{Holistic Sound}\\\textbf{Localization}\\\textbf{(Task 1)}\end{tabular} & 
\begin{tabular}[c]{@{}c@{}}\textbf{Selective Sound}\\\textbf{Localization}\\\textbf{(Task 2)}\end{tabular} & 
\begin{tabular}[c]{@{}c@{}}\textbf{Holistic}\\\textbf{Captioning}\\\textbf{(Task 3)}\end{tabular} & 
\begin{tabular}[c]{@{}c@{}}\textbf{Role-Aware}\\\textbf{Transcription}\\\textbf{(Task 4)}\end{tabular} & 
\begin{tabular}[c]{@{}c@{}}\textbf{Targeted Wearer}\\\textbf{Speech}\\\textbf{Transcription}\\\textbf{(Task 5)}\end{tabular} & 
\begin{tabular}[c]{@{}c@{}}\textbf{Targeted External}\\\textbf{Speech}\\\textbf{Transcription}\\\textbf{(Task 6)}\end{tabular} \\
\midrule
External Non-Speech Sounds & \cmark & \cmark & \cmark & \xmark & \xmark & \xmark \\
Wearer Speech & \xmark & \xmark & \cmark & \cmark & \cmark & \xmark \\
Bystander Speech & \cmark & \xmark & \cmark & \cmark & \xmark & \cmark \\
\bottomrule
\end{tabular}
}
\end{table*}

All tasks use a JSON \texttt{sources} list enclosed by \texttt{<answer>} and \texttt{</answer>}. Every source has a type, which is a description of the sound source (e.g., wearer speech, cat meow), and temporal boundaries. External sources have a relative spatial position specified in spherical coordinates relative to the wearer. The azimuth, elevation, and distance of these coordinates are provided along with the categorical direction and distance. Speech sources contain transcripts. Continuous values are rounded to two decimals. Categorical directions use ten labels spanning primary horizontal, diagonal, and vertical directions (Appendix~\ref{sec:direction_bins}); distance labels are \textit{close} ($<1$ m), \textit{medium range} ($1$--$3$ m), and \textit{far} ($>3$ m). The spatial information for wearer speech is null, whereas bystander speech always has spatial metadata. The examples below are pretty-printed for readability but serialized compactly during training.

\noindent\textbf{Examples.}

\noindent\textbf{Task 1: Holistic Sound Localization.}
This task returns every external sound source with its location and time interval. Transcripts are omitted.

\noindent\textbf{Prompt:} Locate all external sound sources present in the environment.

\begin{footnotesize}
\begin{verbatim}
<answer>
{"sources":[
    {"source_type":"brass_instrument",
     "start_time_s":2.46,"end_time_s":3.05,
     "spatial_info":{"location_description":"right",
         "distance_description":"medium range",
         "azimuth_deg":204.23,"elevation_deg":6.07,
         "distance_m":1.3},
     "transcript":null}
]}
</answer>
\end{verbatim}
\end{footnotesize}

\noindent\textbf{Task 2: Selective Sound Localization.}
This task returns all instances of a queried external non-speech class.

\noindent\textbf{Prompt:} Where is the siren located?

\begin{footnotesize}
\begin{verbatim}
<answer>
{"sources":[
    {"source_type":"siren",
     "start_time_s":2.51,"end_time_s":3.5,
     "spatial_info":{"location_description":"left",
         "distance_description":"far",
         "azimuth_deg":128.4,"elevation_deg":1.17,
         "distance_m":4.53},
     "transcript":null}
]}
</answer>
\end{verbatim}
\end{footnotesize}

\noindent\textbf{Task 3: Holistic Captioning.}
This task returns all wearer speech, bystander speech, and non-speech events, including the attributes applicable to each source type.

\noindent\textbf{Prompt:} Locate, transcribe, and list all audio events and speech in this clip.

\begin{footnotesize}
\begin{verbatim}
<answer>
{"sources":[
    {"source_type":"wearer_speech",
     "start_time_s":7.66,"end_time_s":8.26,
     "spatial_info":null,"transcript":"What is it?\n"},
    {"source_type":"siren",
     "start_time_s":6.05,"end_time_s":7.36,
     "spatial_info":{"location_description":"front",
         "distance_description":"far",
         "azimuth_deg":341.2,"elevation_deg":-30.07,
         "distance_m":7.04},
     "transcript":null}
]}
</answer>
\end{verbatim}
\end{footnotesize}

\noindent\textbf{Task 4: Role-Aware Transcription.}
This task retains only wearer and bystander speech. Bystander speech has spatial metadata, while wearer speech doesn't.

\noindent\textbf{Prompt:} Provide a structured transcription of all speech in this audio, identifying speaker roles and locations.

\begin{footnotesize}
\begin{verbatim}
<answer>
{"sources":[
    {"source_type":"wearer_speech",
     "start_time_s":2.26,"end_time_s":2.83,
     "spatial_info":null,"transcript":"Hi, how are you?\n"},
    {"source_type":"bystander_speech",
     "start_time_s":4.74,"end_time_s":6.9,
     "spatial_info":{"location_description":"front",
         "distance_description":"far",
         "azimuth_deg":8.02,"elevation_deg":-34.83,
         "distance_m":4.77},
     "transcript":"Great weather today!\n"}
]}
</answer>
\end{verbatim}
\end{footnotesize}

\noindent\textbf{Task 5: Targeted Wearer Transcription.}
This task returns only speech produced by the device wearer.

\noindent\textbf{Prompt:} Transcribe all speech spoken by the wearer.

\begin{footnotesize}
\begin{verbatim}
<answer>
{"sources":[
    {"source_type":"wearer_speech",
     "start_time_s":5.04,"end_time_s":5.74,
     "spatial_info":null,
     "transcript":"How do you do?\n"}
]}
</answer>
\end{verbatim}
\end{footnotesize}

\noindent\textbf{Task 6: Targeted External Transcription.}
This task returns bystander speech from the queried categorical direction.

\noindent\textbf{Prompt:} Transcribe the speech segments coming from left.

\begin{footnotesize}
\begin{verbatim}
<answer>
{"sources":[
    {"source_type":"bystander_speech",
     "start_time_s":2.67,"end_time_s":3.32,
     "spatial_info":{"location_description":"left",
         "distance_description":"far",
         "azimuth_deg":96.33,"elevation_deg":10.2,
         "distance_m":5.18},
     "transcript":"\"That's a nice car.\n"}
]}
</answer>
\end{verbatim}
\end{footnotesize}

For negative queries in Tasks 2 and 6, the same schema represents the absence of a matching source without a separate natural-language denial:
\begin{footnotesize}
\begin{verbatim}
<answer>{"sources":[]}</answer>
\end{verbatim}
\end{footnotesize}

\section{\method{} Evaluation Protocol and Metrics}
\label{sec:full_metrics_suite}

This appendix specifies the complete metric suite used for the six tasks. All models are scored with the same parser, source-assignment procedure, and aggregation rules. The task-specific results for these metrics are reported in Appendix~\ref{appendix:evaluation_set_complete}.

\subsection{Common Evaluation Protocol}

\noindent\textbf{Spatial-temporal source assignment.}
Both predicted and ground truth outputs can contain multiple sound sources. To compute metrics like localization error or word error rate, we must first establish a correspondence between them. We match these external sources using the Hungarian algorithm. For reference source $i$ and predicted source $j$, the assignment cost is

\begin{equation}
C_{ij}=\alpha\,d_{\mathrm{arc}}(i,j)+\beta\left(1-\operatorname{tIoU}(i,j)\right),
\qquad \alpha=\beta=0.5,
\label{eq:eval_matching_cost}
\end{equation}

where $d_{\mathrm{arc}}\in[0,1]$ is the great-circle distance between the two directions, normalized by $\pi$, and

\begin{equation}
\operatorname{tIoU}(i,j)=
\frac{\max\!\left(0,\min(e_i,e_j)-\max(s_i,s_j)\right)}
{(e_i-s_i)+(e_j-s_j)-\max\!\left(0,\min(e_i,e_j)-\max(s_i,s_j)\right)}
\label{eq:eval_temporal_iou}
\end{equation}

is the temporal intersection over union for intervals $[s_i,e_i]$ and $[s_j,e_j]$, where $s$ and $e$ denote the start and end times, respectively.

\noindent\textbf{Parsing and format adherence.}
Predictions and references follow the structured JSON contract described in Section~\ref{subsubsec:spatial_audio_sft} (see Appendix~\ref{sec:qa_examples} for examples). A prediction is parsed only if it is valid JSON and satisfies the schema invariants: external sources contain spatial information, speech sources contain transcripts, non-speech sources do not contain transcripts, wearer speech has no spatial information, and every start time precedes its end time. A malformed or schema-invalid prediction is recorded as a format failure and treated as an empty prediction. If $N$ examples are assessed and $N_{\mathrm{valid}}$ are parsed, format adherence is
\begin{equation*}
\text{Format Adherence\,(\%)} = 100\frac{N_{\mathrm{valid}}}{N}.
\end{equation*}

\noindent\textbf{External-source assignment.}
For Tasks~1--4 and 6, predicted and reference external (non-wearer) sources are matched with the Hungarian algorithm (\citep{kuhn1955hungarian}) using the spatial-temporal cost in Equation~\ref{eq:eval_matching_cost}. Assignments whose cost exceeds $0.7$ are rejected. We selected this threshold through empirical manual inspection of more than 20 training examples, fixed it before evaluation, and did not conduct a broader threshold-sensitivity analysis. Each unmatched prediction is a hallucinated source and each unmatched reference is a missing source. The reported hallucination and missing-source rates are the percentages of parsed examples containing at least one such source:
\begin{equation*}
\text{Hallucination Rate\,(\%)} = 100\frac{N_{\mathrm{hallucination}}}{N_{\mathrm{valid}}},
\qquad
\text{Missing Source Rate\,(\%)} = 100\frac{N_{\mathrm{missing}}}{N_{\mathrm{valid}}}.
\end{equation*}
Localization, temporal, source-type, and transcription errors are computed only for accepted matches; these conditional errors must be interpreted alongside the hallucination and missing-source rates.

\noindent\textbf{Wearer-source assignment.}
Wearer speech has no external location and is therefore excluded from Hungarian matching. In Tasks~3, 4, and~5, the predicted and reference wearer sources are matched directly when both are present. If either is missing, the discrepancy is quantified using the wearer detection metrics detailed in the next section. Importantly, a missing or hallucinated wearer source does not contribute to the hallucinated-source and missing-source metrics defined above. This separation ensures the metrics provide sufficient granularity to distinguish between wearer and external source errors.

\subsection{Metric Definitions}

\noindent\textbf{Spatial and semantic metrics.}
For an accepted source pair with reference distance, elevation and azimuth $(r,\theta,\phi)$ and prediction $(\hat{r},\hat{\theta},\hat{\phi})$, the azimuth error accounts for circular wrap-around,
\begin{equation*}
\text{Azimuth Error}\,(^\circ) = \min\!\left(|\hat{\phi}-\phi|, 360^\circ-|\hat{\phi}-\phi|\right),
\end{equation*}
while elevation and distance errors are $|\hat{\theta}-\theta|$ and $|\hat{r}-r|$, respectively. Percentage distance error is
\begin{equation*}
\text{Distance Error\,(\%)} = 100\frac{|\hat{r}-r|}{\max(r, 10^{-5})}.
\end{equation*}
Azimuth MAE, elevation MAE, distance MAE, and percentage distance error are pooled over matched pairs of external sources across all examples.

In addition to continuous-valued estimates, we discretize the external source directions and distances into bins. The bin definitions are provided in Appendix~\ref{sec:direction_bins}. Distance bins are \emph{close} ($r < 1$ m), \emph{medium range} ($1 \leq r \leq 3$ m), and \emph{far} ($r > 3$ m). Location-bin accuracy and distance-bin accuracy are the exact-match rates, pooled via micro-averaging over matched pairs of external sources across the dataset.

To compute source-type accuracy, we normalize the predicted and reference sound descriptors (e.g., \emph{dog barking}, \emph{speech}) by converting the strings to lowercase and stripping leading English articles. We then map these intermediate strings to canonical labels using a predefined synonym dictionary (e.g., mapping \emph{door knock} to \emph{knock}). Source-type accuracy is the fraction of matched sources whose canonicalized labels agree, pooled via micro-averaging over matched pairs of external sources across the dataset.

\noindent\textbf{Temporal metrics.}
Temporal errors are measured using start-time Mean Absolute Error (MAE), end-time MAE, and Temporal Intersection over Union (IoU). Let $s$ and $e$ denote the reference start and end times, and let $\hat{s}$ and $\hat{e}$ denote the corresponding predicted start and end times. Start-time MAE and end-time MAE represent the absolute differences $|\hat{s}-s|$ and $|\hat{e}-e|$. Temporal IoU evaluates the overlap of the temporal boundaries (Equation~\ref{eq:eval_temporal_iou}). These metrics are computed for accepted source pairs and pooled via micro-averaging across the entire dataset. Unmatched or missing sources do not contribute to temporal errors. The specific source types comprising the matched pool depend on the task definition. For example, holistic captioning (Task~3) and role-aware transcription (Task~4) compute errors over both wearer and external sources. Holistic sound localization (Task~1) evaluates all external sources, and targeted wearer speech transcription (Task~5) evaluates only wearer speech. Selective sound localization (Task~2) and targeted external speech transcription (Task~6) restrict evaluation to external sources satisfying the queried sound class or location condition.

\noindent\textbf{Transcription.}
Transcription accuracy is measured using Word Error Rate (WER), capped at $100\%$. Predicted and reference transcripts are standardized by removing filler words, converting to lowercase, replacing punctuation with spaces, and collapsing extraneous whitespace. WER is individually computed over matched source pairs and then micro-averaged across the dataset. Holistic sound localization (Task~1) and selective sound localization (Task~2) do not evaluate transcription.

\noindent\textbf{Detection metrics.}
Selective Sound Localization (Task~2) and Targeted External Speech Transcription (Task~6) evaluate query detection at the source level. Accepted source assignments are true positives (TP), unmatched predicted sources are false positives (FP), unmatched references are false negatives (FN), and an example with an absent target source and an empty prediction contributes one true negative (TN). Holistic Captioning (Task~3), Role-Aware Transcription (Task~4), and Targeted Wearer Speech Transcription (Task~5) instead evaluate wearer detection at the example level: TP and TN indicate agreement that wearer speech is present or absent, while FP and FN denote a hallucinated or omitted wearer source. Both settings report
\begin{equation*}
\mathrm{Accuracy}=100\frac{\mathrm{TP}+\mathrm{TN}}{\mathrm{TP}+\mathrm{TN}+\mathrm{FP}+\mathrm{FN}},\quad
\mathrm{Precision}=100\frac{\mathrm{TP}}{\mathrm{TP}+\mathrm{FP}},\quad
\mathrm{Recall}=100\frac{\mathrm{TP}}{\mathrm{TP}+\mathrm{FN}},
\end{equation*}
and $\mathrm{F1}=2PR/(P+R)$, with a zero value whenever the corresponding denominator is zero. The global counts for true positives, false positives, false negatives, and true negatives are accumulated across the entire dataset, after which the formulas are applied directly over these aggregate integer pools to yield the final scores.

\subsection{Metrics by Task}

\providecommand{\cmark}{\ding{51}}
\providecommand{\xmark}{\ding{55}}

\begin{table*}[t]
\centering
\scriptsize
\caption{Applicability of metric groups to the six evaluation tasks. ``External'' refers to sources other than the wearer.}
\label{tab:metric_suite_matrix}
\setlength{\tabcolsep}{4pt}
\resizebox{\linewidth}{!}{
\begin{tabular}{@{}lcccccc@{}}
\toprule
\textbf{Metric group} & 
\begin{tabular}[c]{@{}c@{}}\textbf{Holistic Sound}\\\textbf{Localization}\\\textbf{(Task 1)}\end{tabular} & 
\begin{tabular}[c]{@{}c@{}}\textbf{Selective Sound}\\\textbf{Localization}\\\textbf{(Task 2)}\end{tabular} & 
\begin{tabular}[c]{@{}c@{}}\textbf{Holistic}\\\textbf{Captioning}\\\textbf{(Task 3)}\end{tabular} & 
\begin{tabular}[c]{@{}c@{}}\textbf{Role-Aware}\\\textbf{Transcription}\\\textbf{(Task 4)}\end{tabular} & 
\begin{tabular}[c]{@{}c@{}}\textbf{Targeted Wearer}\\\textbf{Speech}\\\textbf{Transcription}\\\textbf{(Task 5)}\end{tabular} & 
\begin{tabular}[c]{@{}c@{}}\textbf{Targeted External}\\\textbf{Speech}\\\textbf{Transcription}\\\textbf{(Task 6)}\end{tabular} \\
\midrule
Format Adherence & \cmark & \cmark & \cmark & \cmark & \cmark & \cmark \\
Hallucinated/Missing External Sources & \cmark & \cmark & \cmark & \cmark & \xmark & \cmark \\
Source-Type Accuracy & \cmark & \xmark & \cmark & \xmark & \xmark & \xmark \\
Spatial Errors and Bin Accuracies & \cmark & \cmark & \cmark & \cmark & \xmark & \cmark \\
Temporal Errors and IoU & \cmark & \cmark & \cmark & \cmark & \cmark & \cmark \\
Speech Transcription (WER) & \xmark & \xmark & \cmark & \cmark & \cmark & \cmark \\
Wearer Detection & \xmark & \xmark & \cmark & \cmark & \cmark & \xmark \\
Query Detection & \xmark & \cmark & \xmark & \xmark & \xmark & \cmark \\
\bottomrule
\end{tabular}
}
\end{table*}

Table~\ref{tab:metric_suite_matrix} summarizes the applicability of each metric group across the six evaluation tasks. The specific evaluation protocol and metrics computed for each task are detailed below.

\noindent\textbf{Task 1: Holistic Sound Localization.}
All predicted and reference external sources participate in Hungarian assignment. We report format adherence; external-source hallucination and missing-source rates; source-type accuracy; azimuth, elevation, distance, and percentage-distance errors; location- and distance-bin accuracies; start- and end-time errors; and temporal IoU.

\noindent\textbf{Task 2: Selective Sound Localization.}
Predictions and references are restricted to the queried sound class before assignment. We report format adherence; external-source hallucination and missing-source rates; azimuth, elevation, distance, and percentage-distance errors; location- and distance-bin accuracies; start- and end-time errors; temporal IoU; and source-level query-detection accuracy, precision, recall, and F1. Empty predictions for absent queried classes are true negatives.

\noindent\textbf{Task 3: Holistic Captioning.}
Wearer speech is matched directly when present in both the prediction and reference, while all external speech and non-speech sources are matched using the Hungarian algorithm. We report format adherence; external-source hallucination and missing-source rates; source-type accuracy on matched external sources; spatial and bin metrics on matched external sources; temporal metrics on all matched sources; WER on matched speech; and example-level wearer-detection accuracy, precision, recall, and F1.

\noindent\textbf{Task 4: Role-Aware Transcription.}
Only speech sources are evaluated. Wearer speech is matched directly, while bystander speech is matched using the Hungarian algorithm. We report format adherence; bystander-speech hallucination and missing-source rates; spatial and bin metrics on matched bystander speech; temporal metrics and WER on all matched speech; and example-level wearer-detection accuracy, precision, recall, and F1.

\noindent\textbf{Task 5: Targeted Wearer Speech Transcription.}
Only wearer speech is evaluated. We report format adherence; start- and end-time errors; temporal IoU; wearer WER; and example-level wearer-detection accuracy, precision, recall, and F1.

\noindent\textbf{Task 6: Targeted External Speech Transcription.}
Predictions and references are restricted to bystander speech in the queried direction before Hungarian assignment. We report format adherence; external-speech hallucination and missing-source rates; azimuth, elevation, distance, and percentage-distance errors; location- and distance-bin accuracies; start- and end-time errors; temporal IoU; WER; and source-level query-detection accuracy, precision, recall, and F1. Empty predictions for unoccupied queried directions are true negatives.

\section{Cohort Analysis}
\label{appendix:evaluation_set_cohorts}

We analyze performance on the evaluation set across four distinct cohort axes: \textit{total source count}, \textit{speaker count}, \textit{microphone count}, and \textit{source overlap percentage}. The \textit{total source count} cohort groups scenes by the total number of active speech and non-speech sources (inclusive of wearer speech, bystander speech, and non-speech sounds) into 1, 2, or 3+ sources. Similarly, the \textit{speaker count} cohort isolates speech sources by counting both wearer speech and bystander speech, grouped into 1, 2, or 3+ speakers. The \textit{microphone count} cohort stratifies scenes by the 4--9 microphones present in their simulated array.

To evaluate performance under varying degrees of acoustic collision, we introduce the \textit{source overlap percentage} cohort. For an acoustic scene with active sound source intervals $I_i = [s_i, e_i)$, we calculate the total active duration $T_{\mathrm{active}}$, the overlapping active duration $T_{\mathrm{overlap}}$, and the relative overlap percentage $R_{\mathrm{overlap}}$ as:
\begin{equation}
T_{\mathrm{active}}=\mu\!\left(\bigcup_i I_i\right),\qquad
T_{\mathrm{overlap}}=\mu\!\left(\left\{t:\sum_i \mathbf{1}[t\in I_i]\geq 2\right\}\right),
\qquad
R_{\mathrm{overlap}}=100\cdot\frac{T_{\mathrm{overlap}}}{T_{\mathrm{active}}},
\label{eq:evaluation_set_overlap_cohort}
\end{equation}
where $\mu(\cdot)$ denotes cumulative time duration, and $\mathbf{1}[\cdot]$ is an indicator function that evaluates to 1 when source $i$ is active at time $t$ and 0 otherwise. Here, $T_{\mathrm{active}}$ represents the total duration where at least one sound source is active, while $T_{\mathrm{overlap}}$ represents the cumulative duration where two or more sources are concurrently active. The scalar $R_{\mathrm{overlap}}$ measures the overlapping duration as a percentage of the total active duration. We partition $R_{\mathrm{overlap}}$ into five discrete cohorts: $\{0\}$, $(0,25]$, $(25,50]$, $(50,75]$, and $(75,100]$ percent. Invalid or zero-duration intervals are omitted when computing these durations.

\begin{figure*}[t]
    \centering
    \includegraphics[width=\textwidth]{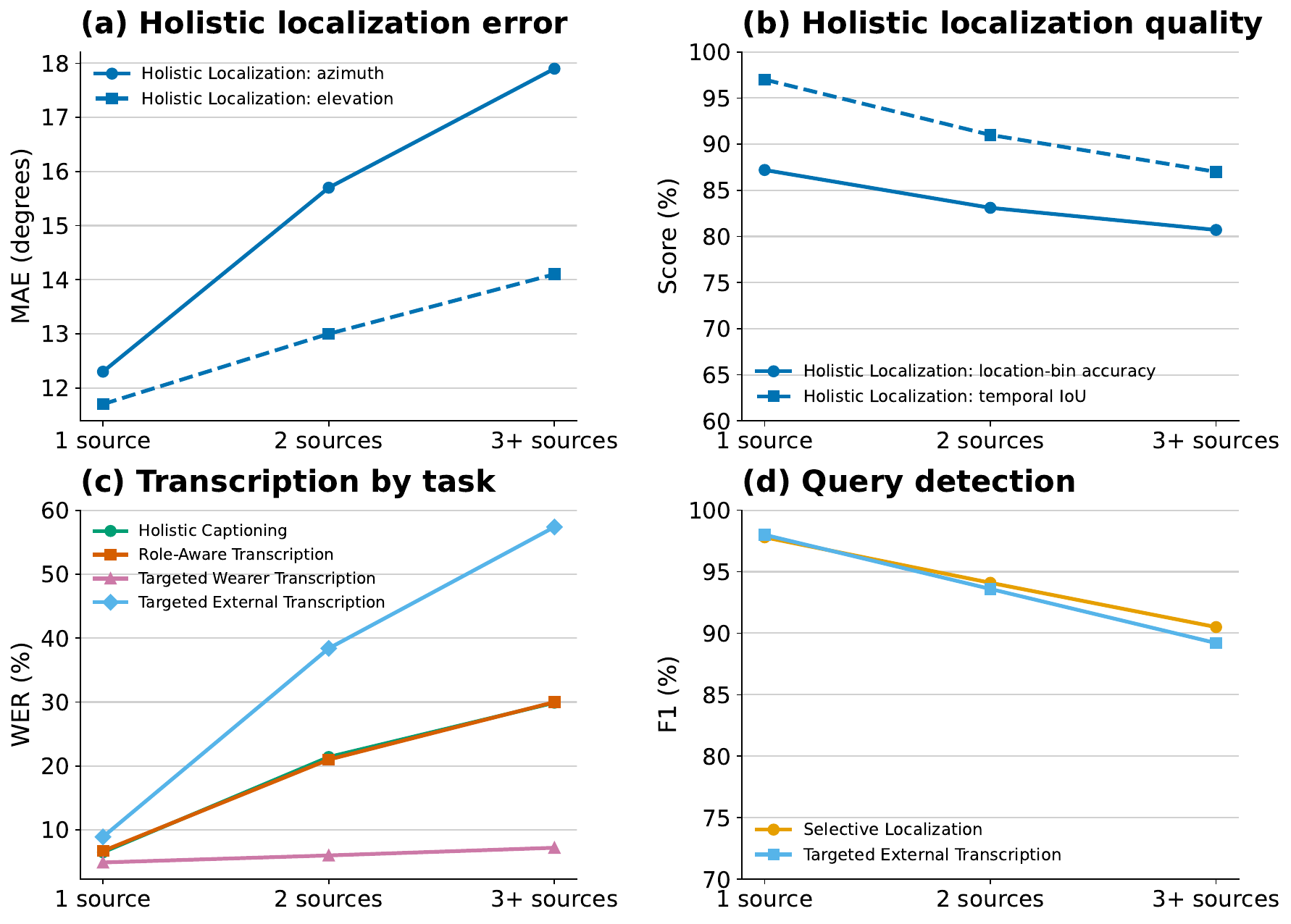}
    \caption{\small\textbf{Results segmented by total source count.} (a)~\textbf{Holistic Localization Error:} Task~1 azimuth and elevation MAEs across total-source-count cohorts. (b)~\textbf{Holistic Localization Quality:} Task~1 location-bin accuracy and temporal IoU across the same cohorts. (c)~\textbf{Transcription by Task:} WER for Holistic Captioning, Role-Aware Transcription, Targeted Wearer Transcription, and Targeted External Transcription. (d)~\textbf{Query Detection:} Query F1 for Selective Localization and Targeted External Transcription.}
    \label{fig:evaluation_source_density}
\end{figure*}

Figure~\ref{fig:evaluation_source_density} demonstrates performance trends as the number of sound sources increases. Specifically, spatial localization quality and temporal precision degrade with higher source counts (Figures~\ref{fig:evaluation_source_density}a and~\ref{fig:evaluation_source_density}b). For Holistic Localization, azimuth MAE increases as more sound sources are added (Figure~\ref{fig:evaluation_source_density}a), while location-bin accuracy and temporal IoU both decrease (Figure~\ref{fig:evaluation_source_density}b). Similarly, transcription quality degrades substantially for external sound sources as source count increases, whereas wearer speech transcription is not significantly affected, likely due to the favorable microphone proximity of the wearer (Figure~\ref{fig:evaluation_source_density}c). Word error rates rise noticeably across Holistic Captioning, Role-Aware Transcription, and Targeted External Transcription, while Targeted Wearer Transcription WER remains comparatively stable. Finally, query detection reliability also degrades with higher source counts (Figure~\ref{fig:evaluation_source_density}d), with query F1 scores decreasing for both Selective Localization and Targeted External Transcription.

\begin{figure*}[t]
    \centering
    \includegraphics[width=\textwidth]{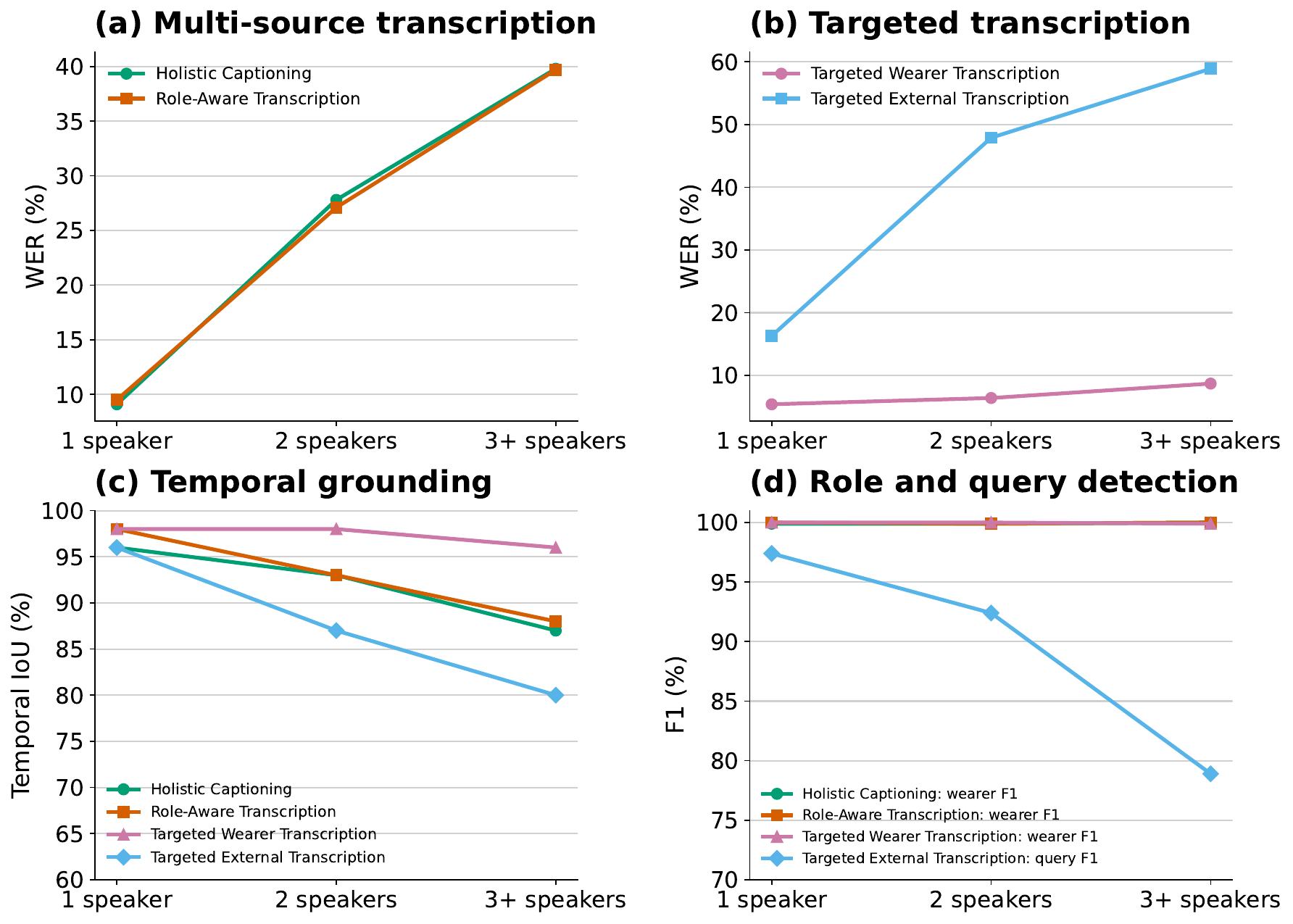}
    \caption{\small\textbf{Results segmented by speaker count.} Speaker count includes wearer- and bystander-speech sources in the complete scene. (a)~\textbf{Multi-Source Transcription:} WER for Holistic Captioning and Role-Aware Transcription across speaker-count cohorts. (b)~\textbf{Targeted Transcription:} WER for Targeted Wearer Transcription and Targeted External Transcription. (c)~\textbf{Temporal Grounding:} Temporal IoU for the four transcription tasks. (d)~\textbf{Role and Query Detection:} Wearer F1 for Holistic Captioning, Role-Aware Transcription, and Targeted Wearer Transcription, together with query F1 for Targeted External Transcription. The three wearer-F1 traces overlap almost completely near $100\%$, so they partially obscure one another.}
    \label{fig:evaluation_speaker_density}
\end{figure*}

Figure~\ref{fig:evaluation_speaker_density} shows similar trends to the source-count analysis: as the number of speech sources increases, acoustic scene complexity grows, making external speech transcription and direction-targeted transcription of external sources substantially more challenging. As shown in Figures~\ref{fig:evaluation_speaker_density}a and~\ref{fig:evaluation_speaker_density}b, word error rates increase for Holistic Captioning, Role-Aware Transcription, and Targeted External Transcription as external speech becomes harder to transcribe. In contrast, Targeted Wearer Speech Transcription is not as substantially impacted (Figure~\ref{fig:evaluation_speaker_density}b). Similarly, temporal metrics degrade for tasks involving external sources as isolating those sources becomes harder, whereas temporal performance for Targeted Wearer Transcription does not degrade as speaker count increases (Figure~\ref{fig:evaluation_speaker_density}c). Finally, Figure~\ref{fig:evaluation_speaker_density}d indicates that query detection becomes less reliable for Targeted External Speech Transcription as the number of speakers increases, whereas wearer speech detection remains robust. Note that the three-or-more-speaker cohort contains only 546--586 evaluated examples per task, making its estimates less stable than those for the one- and two-speaker cohorts.

\begin{figure*}[t]
    \centering
    \includegraphics[width=\textwidth]{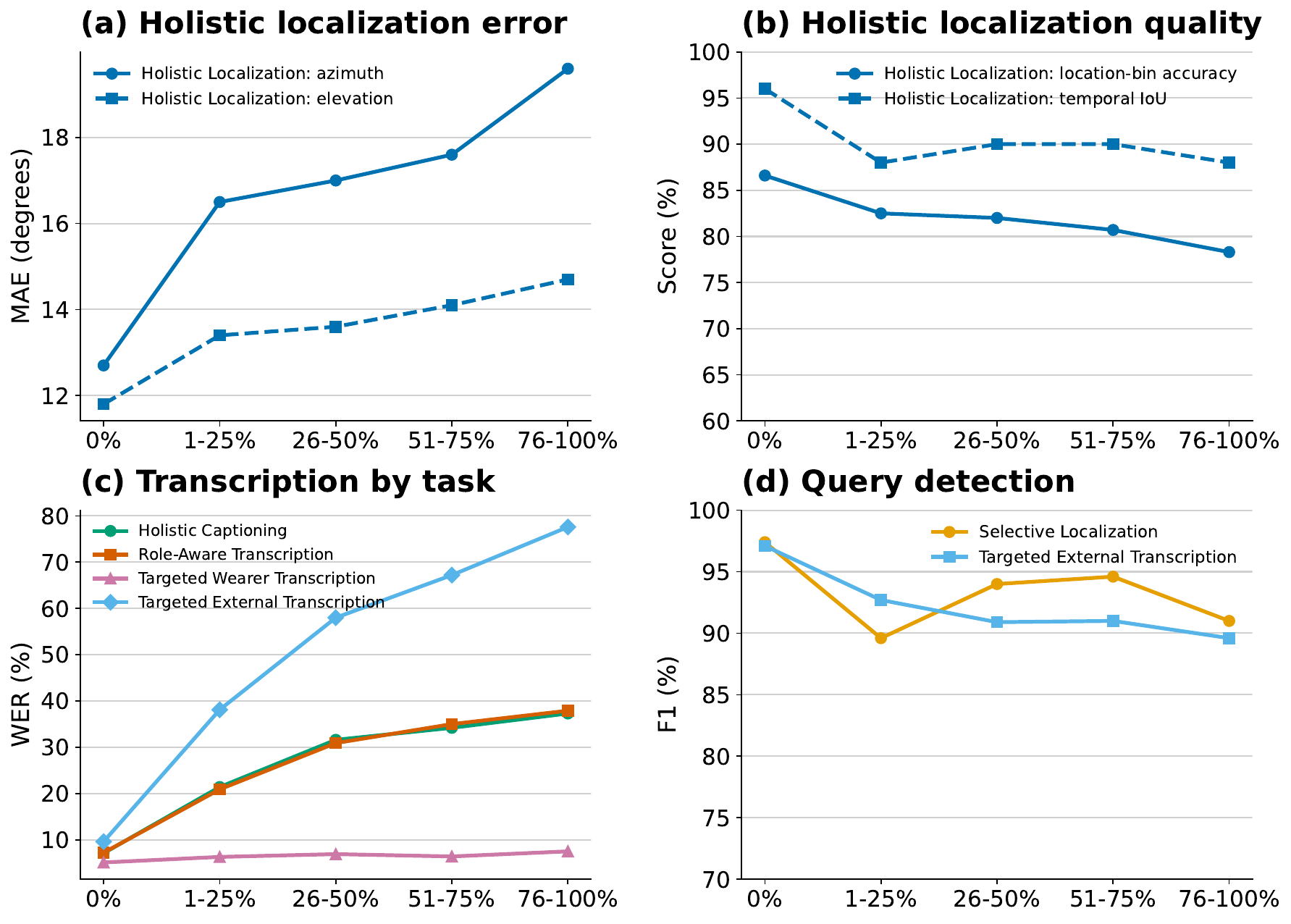}
    \caption{\small\textbf{Results segmented by source-overlap percentage.} Overlap includes all speech and non-speech sources and is normalized by the union of active source intervals as defined in Equation~\ref{eq:evaluation_set_overlap_cohort}. (a)~\textbf{Holistic Localization Error:} Task~1 azimuth and elevation MAEs across overlap cohorts. (b)~\textbf{Holistic Localization Quality:} Task~1 location-bin accuracy and temporal IoU. (c)~\textbf{Transcription by Task:} WER for Holistic Captioning, Role-Aware Transcription, Targeted Wearer Transcription, and Targeted External Transcription. (d)~\textbf{Query Detection:} Query F1 for Selective Localization and Targeted External Transcription.}
    \label{fig:evaluation_overlap_percentage}
\end{figure*}

Figure~\ref{fig:evaluation_overlap_percentage} shows similar trends to the source-count analysis when segmenting results by source overlap percentage. As source overlap increases, azimuth and elevation localization errors grow while location-bin accuracy degrades (Figures~\ref{fig:evaluation_overlap_percentage}a and~\ref{fig:evaluation_overlap_percentage}b). Temporal IoU also degrades with higher overlap (Figure~\ref{fig:evaluation_overlap_percentage}b). As shown in Figure~\ref{fig:evaluation_overlap_percentage}c, word error rates increase substantially with higher overlap percentage for tasks that require transcribing external sources, whereas Targeted Wearer Speech Transcription is not as significantly impacted. Finally, Figure~\ref{fig:evaluation_overlap_percentage}d shows that query detection for external sources degrades as source overlap increases. Note that the highest-overlap cohort contains only 508--836 evaluated examples per task, making its estimates less stable than those for lower overlap cohorts.

\begin{figure*}[t]
    \centering
    \includegraphics[width=\textwidth]{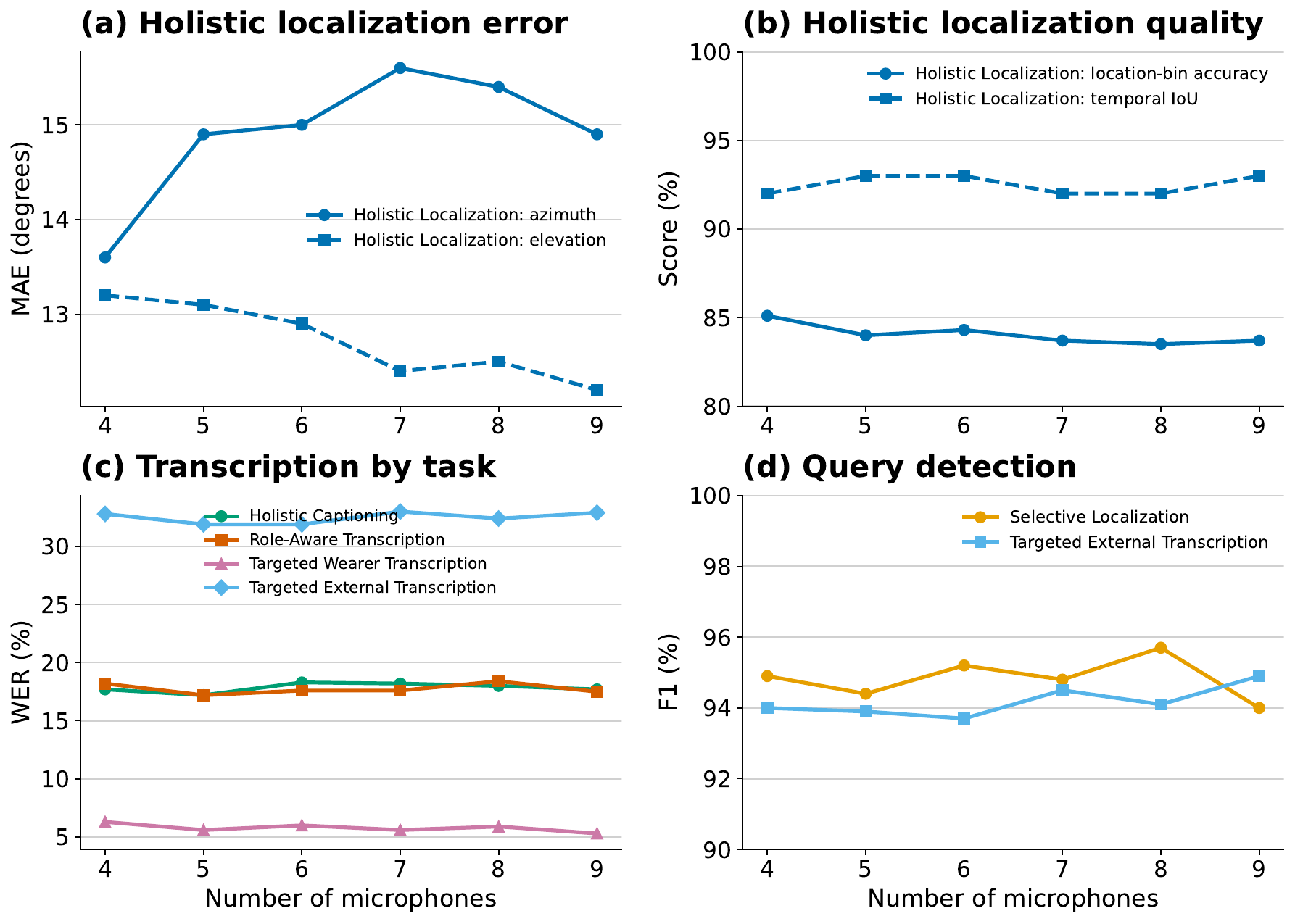}
    \caption{\small\textbf{Results segmented by array microphone count.} A single \method{} checkpoint is evaluated without array-specific retraining across scenes synthesized with 4--9 microphones. (a)~\textbf{Holistic Localization Error:} Task~1 azimuth and elevation MAEs across microphone counts. (b)~\textbf{Holistic Localization Quality:} Task~1 location-bin accuracy and temporal IoU. (c)~\textbf{Transcription by Task:} WER for Holistic Captioning, Role-Aware Transcription, Targeted Wearer Transcription, and Targeted External Transcription. (d)~\textbf{Query Detection:} Query F1 for Selective Localization and Targeted External Transcription.}
    \label{fig:evaluation_microphone_count}
\end{figure*}

Figure~\ref{fig:evaluation_microphone_count} presents performance results segmented by the number of microphones in the array. As shown in Figures~\ref{fig:evaluation_microphone_count}a and~\ref{fig:evaluation_microphone_count}b, localization errors and temporal IoU are not significantly affected by microphone count; similarly, transcription WER (Figure~\ref{fig:evaluation_microphone_count}c) and query detection reliability (Figure~\ref{fig:evaluation_microphone_count}d) remain stable across array sizes. Together with evaluation across 1,211 configurations, this stability shows that one checkpoint accommodates varied microphone counts and positions within the tested smart-glasses design space. It does not demonstrate generalization to unseen topologies. Moreover, the result contrasts with PhaseCoder \citep{dementyev2026phasecoder}, which found that increasing microphone count generally reduces localization error. PhaseCoder evaluated fully randomized layouts with high spatial variability, whereas our configurations are subsets of microphone positions on only two fixed glasses frames. Their $z$-axis displacement is inherently constrained, and the candidate locations are not positioned completely at random. This restricted spatial spread may obscure performance trends across microphone counts that would emerge with spherical, circular, linear, fully randomized, or non-glasses arrays.

\section{Categorical Direction Bins}
\label{sec:direction_bins}

We use categorical direction labels to provide a coarse, human-readable description of each external source's direction. The source direction is mapped to one of six location bins whose elevation ($\theta$) and azimuth ($\phi$) ranges are specified in Table~\ref{tab:direction_bins}. Labels are serialized as lowercase literals without an article.

\begin{table}[h]
\centering
\small
\caption{Categorical direction bins and their corresponding elevation ($\theta$) and azimuth ($\phi$) ranges in degrees.}
\label{tab:direction_bins}
\begin{tabular}{@{}lll@{}}
\toprule
\textbf{Location Bin} & \textbf{Elevation Range ($\theta$)} & \textbf{Azimuth Range ($\phi$)} \\
\midrule
\textit{above} & $(80^\circ, 90^\circ]$ & $[0^\circ, 360^\circ)$ \\
\textit{below} & $[-90^\circ, -80^\circ)$ & $[0^\circ, 360^\circ)$ \\
\textit{front} & $[-80^\circ, 80^\circ]$ & $(337.5^\circ, 360^\circ) \cup [0^\circ, 22.5^\circ]$ \\
\textit{left} & $[-80^\circ, 80^\circ]$ & $(22.5^\circ, 157.5^\circ]$ \\
\textit{rear} & $[-80^\circ, 80^\circ]$ & $(157.5^\circ, 202.5^\circ]$ \\
\textit{right} & $[-80^\circ, 80^\circ]$ & $(202.5^\circ, 337.5^\circ]$ \\
\bottomrule
\end{tabular}
\end{table}

\section{Training Stages for \method{}}
\label{appendix:training_stages}
\method{} is trained using a two-stage curriculum on our synthetic training and validation partitions, without mixing external benchmark data.

\noindent\textbf{Stage 1: Single-Source Projector Alignment.}
We initialize from \basemodel{} and freeze the LLM, its native audio tower, \samethod{}, the token embeddings, and the language-model head. Only the randomly initialized spatial projector is trainable. To provide a gradual introduction to the training curriculum with a relatively easier task, we train exclusively on Holistic Sound Localization (Task~1) and filter the training data to include exactly one sound source per recording. For each recording, we construct one prompt-response pair. We train for three epochs with a learning rate of $5\times10^{-4}$, spatial and temporal token weights set to 5, a per-device batch size of 4, and two gradient-accumulation steps. Across eight GPUs, the effective global batch size is 64.

\noindent\textbf{Stage 2: Joint Holistic SFT.}
We warm-start from the best Stage~1 checkpoint and introduce LoRA adapters on every LLM attention and feed-forward projection while continuing to train the spatial projector. The adapters use rank 128, $\alpha=256$, and dropout 0.05; the base LLM, token embeddings, language-model head, native audio tower, and \samethod{} remain frozen. We uniformly sample examples across all six tasks, with scenes containing one to four sound sources. For each recording, we construct up to two prompt-response pairs from different tasks, and generate negative queries for Selective Sound Localization (Task~2) and Targeted External Transcription (Task~6) with a probability of 0.3. We train for three epochs with a learning rate of $1\times10^{-4}$, spatial and temporal token weights set to 5, a per-device batch size of 2, and four gradient-accumulation steps, preserving an effective global batch size of 64. Both stages use a 2048-token context, AdamW optimizer, and cosine decay learning rate schedule with 5\% warmup.

\section{Curriculum Ablation Studies}
\label{appendix:curse_curriculum}
\subsection{\method{} Curriculum Ablation}
We tested whether a finer-grained curriculum helps \method{} by inserting a focused Stage~2a between projector alignment and joint holistic SFT for training \method{}. When training with the additional training stage, we used the same training dataset and evaluation methodology. Stage~2a trains Selective Sound Localization (Task~2), Targeted Wearer Transcription (Task~5), and Targeted External Transcription (Task~6) on scenes with one or two sound sources before Stage~2 introduces all six tasks and more sound sources. Stage~2a uses the same trainable/frozen component split as the final Stage~2 (Stage~2b in the three-stage curriculum): we train the spatial projector and the LoRA adapters on the LLM attention and feed-forward layers, while the base LLM weights, token embeddings, language-model head, native audio tower, and \samethod{} remain frozen. Table~\ref{tab:curriculum_ablation} shows that this additional stage degrades nearly every reported metric.

\begin{table*}[t]
\centering
\small
\caption{Effect of inserting focused Stage~2a between Stage~1 and Stage~2 on the held-out synthetic evaluation set. Arrows indicate the preferred direction for each metric.}
\label{tab:curriculum_ablation}
\resizebox{\linewidth}{!}{
\begin{tabular}{@{}l*{9}{r}@{}}
\toprule
& \multicolumn{6}{c}{\textbf{Task 1: Holistic Sound Localization}} & \multicolumn{3}{c}{\textbf{Task 2: Selective Sound Localization}} \\
\cmidrule(lr){2-7} \cmidrule(lr){8-10}
\textbf{Curriculum} & \textbf{Azim.} $\downarrow$ & \textbf{Elev.} $\downarrow$ & \textbf{Dist.} $\downarrow$ & \textbf{Loc. bin} $\uparrow$ & \textbf{Type acc.} $\uparrow$ & \textbf{tIoU} $\uparrow$ & \textbf{Azim.} $\downarrow$ & \textbf{Elev.} $\downarrow$ & \textbf{Query F1} $\uparrow$ \\
& ($^\circ$) & ($^\circ$) & (m) & (\%) & (\%) & & ($^\circ$) & ($^\circ$) & (\%) \\
\midrule
Stage~1 $\rightarrow$ Stage~2a $\rightarrow$ Stage~2 & 15.5 & 13.1 & 1.20 & \textbf{85.5} & 61.4 & 0.84 & 19.7 & 15.2 & 83.9 \\
Stage~1 $\rightarrow$ Stage~2 (ours) & \textbf{14.9} & \textbf{12.8} & \textbf{1.07} & 84.1 & \textbf{63.7} & \textbf{0.92} & \textbf{17.5} & \textbf{14.8} & \textbf{94.9} \\
\bottomrule
\end{tabular}}

\vspace{0.5em}
\resizebox{\linewidth}{!}{
\begin{tabular}{@{}l*{8}{r}@{}}
\toprule
& \multicolumn{2}{c}{\textbf{Task 3: Holistic Captioning}} & \multicolumn{2}{c}{\textbf{Task 4: Role-Aware Transcription}} & \multicolumn{2}{c}{\textbf{Task 5: Targeted Wearer Transcription}} & \multicolumn{2}{c}{\textbf{Task 6: Targeted External Transcription}} \\
\cmidrule(lr){2-3} \cmidrule(lr){4-5} \cmidrule(lr){6-7} \cmidrule(lr){8-9}
\textbf{Curriculum} & \textbf{WER} $\downarrow$ & \textbf{Wearer F1} $\uparrow$ & \textbf{WER} $\downarrow$ & \textbf{Wearer F1} $\uparrow$ & \textbf{WER} $\downarrow$ & \textbf{Wearer F1} $\uparrow$ & \textbf{WER} $\downarrow$ & \textbf{Query F1} $\uparrow$ \\
\midrule
Stage~1 $\rightarrow$ Stage~2a $\rightarrow$ Stage~2 & 24.0 & 78.4 & 23.4 & 83.1 & 31.6 & 87.7 & 46.9 & 87.9 \\
Stage~1 $\rightarrow$ Stage~2 (ours) & \textbf{17.8} & \textbf{99.9} & \textbf{17.8} & \textbf{100.0} & \textbf{5.8} & \textbf{100.0} & \textbf{32.5} & \textbf{94.1} \\
\bottomrule
\end{tabular}}
\end{table*}

We hypothesize three interacting causes. First, Stage~2a exposes the model only to tasks that involve selecting a specific source and generating a response based on that selected source, namely, Selective Sound Localization (Task~2), Targeted Wearer Transcription (Task~5), and Targeted External Transcription (Task~6), in scenes containing only one or two sound sources. This restricted distribution may specialize the projector and LoRA adapters toward short, query-conditioned outputs before encountering more complex acoustic scenes and having to produce detailed outputs that include all of the sound sources in the scene. Stage~2 may not be able to undo this specialization. Second, because the native audio tower and \samethod{} remain frozen, the Stage~2a updates are concentrated in the spatial projector and LLM LoRA adapters; this may bias the cross-modal alignment toward the restricted selective-task distribution without allowing the audio representations to adapt. Third, adding a curriculum stage makes the curriculum and optimization trajectory more complex, which may actually be detrimental to reaching a better optimum.

\section{Extended \method{} Evaluation Results}
\label{appendix:evaluation_set_complete}

In Table~\ref{tab:main_results_compact}, we present results for a subset of the evaluation metrics specified in Appendix~\ref{sec:full_metrics_suite}. We report the complete set of metrics across all six tasks for the same four models in Tables~\ref{tab:evaluation_set_task1_complete}--\ref{tab:evaluation_set_task6_complete}. 

\begin{table*}[t]
\centering
\scriptsize
\caption{Complete Task~1 (Holistic Sound Localization) results on the evaluation set. $N$ = total evaluated examples; Format = format adherence; Halluc. = external-source hallucination rate; Missing = missing external source rate; Type acc. = source-type accuracy; Azim. = azimuth MAE; Elev. = elevation MAE; Dist. = distance MAE; Dist.\ err. = percentage distance error; Loc.\ bin = location-bin accuracy; Dist.\ bin = distance-bin accuracy; Start = start-time MAE; End = end-time MAE; tIoU = temporal intersection over union.}
\label{tab:evaluation_set_task1_complete}
\resizebox{0.8\linewidth}{!}{
\begin{tabular}{@{}lrrrrr@{}}
\toprule
\textbf{Model} & $\mathbf{N}$ & \textbf{Format} $\uparrow$ & \textbf{Halluc.} $\downarrow$ & \textbf{Missing} $\downarrow$ & \textbf{Type acc.} $\uparrow$ \\
& & (\%) & (\%) & (\%) & (\%) \\
\midrule
\basemodel{} (vanilla) & 26,985 & 100.0 & 4.0 & 94.8 & 0.9 \\
\basemodel{} (fine-tuned) & 26,985 & 100.0 & 6.8 & 10.6 & 62.2 \\
SELDNet $+$ \basemodel{} & 26,985 & 100.0 & 50.3 & 22.6 & 55.6 \\
\rowcolor{methodrow}\method{} \textit{(ours)} & 26,985 & 100.0 & 3.6 & 5.8 & 63.7 \\
\bottomrule
\end{tabular}}

\vspace{0.35em}
\resizebox{\linewidth}{!}{
\begin{tabular}{@{}lrrrrrrrrr@{}}
\toprule
\textbf{Model} & \textbf{Azim.} $\downarrow$ & \textbf{Elev.} $\downarrow$ & \textbf{Dist.} $\downarrow$ & \textbf{Dist. err.} $\downarrow$ & \textbf{Loc. bin} $\uparrow$ & \textbf{Dist. bin} $\uparrow$ & \textbf{Start} $\downarrow$ & \textbf{End} $\downarrow$ & \textbf{tIoU} $\uparrow$ \\
& ($^\circ$) & ($^\circ$) & (m) & (\%) & (\%) & (\%) & (s) & (s) & \\
\midrule
\basemodel{} (vanilla) & 67.9 & 27.7 & 1.88 & 72.1 & 20.1 & 28.9 & 2.59 & 3.87 & 0.37 \\
\basemodel{} (fine-tuned) & 87.5 & 27.7 & 1.28 & 46.1 & 33.1 & 57.4 & 0.22 & 0.48 & 0.91 \\
SELDNet $+$ \basemodel{} & 80.6 & 27.4 & 1.26 & 43.2 & 42.0 & 60.5 & 1.81 & 2.39 & 0.61 \\
\rowcolor{methodrow}\method{} \textit{(ours)} & 14.9 & 12.8 & 1.07 & 37.5 & 84.1 & 67.0 & 0.15 & 0.25 & 0.92 \\
\bottomrule
\end{tabular}}
\end{table*}

\begin{table*}[t]
\centering
\scriptsize
\caption{Complete Task~2 (Selective Sound Localization) results on the evaluation set. $N$ = total evaluated examples; Format = format adherence; Halluc. = external-source hallucination rate; Missing = missing external source rate; Azim. = azimuth MAE; Elev. = elevation MAE; Dist. = distance MAE; Dist.\ err. = percentage distance error; Loc.\ bin = location-bin accuracy; Dist.\ bin = distance-bin accuracy; Start = start-time MAE; End = end-time MAE; tIoU = temporal intersection over union; Detection acc. = source-level query-detection accuracy; Precision = precision; Recall = recall; F1 = F1 score.}
\label{tab:evaluation_set_task2_complete}
\resizebox{0.8\linewidth}{!}{
\begin{tabular}{@{}lrrrr@{}}
\toprule
\textbf{Model} & $\mathbf{N}$ & \textbf{Format} $\uparrow$ & \textbf{Halluc.} $\downarrow$ & \textbf{Missing} $\downarrow$ \\
& & (\%) & (\%) & (\%) \\
\midrule
\basemodel{} (vanilla) & 17,247 & 100.0 & 26.1 & 64.7 \\
\basemodel{} (fine-tuned) & 17,247 & 100.0 & 34.0 & 4.0 \\
SELDNet $+$ \basemodel{} & 17,247 & 100.0 & 21.1 & 13.8 \\
\rowcolor{methodrow}\method{} \textit{(ours)} & 17,247 & 100.0 & 4.5 & 2.9 \\
\bottomrule
\end{tabular}}

\vspace{0.35em}
\resizebox{\linewidth}{!}{
\begin{tabular}{@{}lrrrrrrrrr@{}}
\toprule
\textbf{Model} & \textbf{Azim.} $\downarrow$ & \textbf{Elev.} $\downarrow$ & \textbf{Dist.} $\downarrow$ & \textbf{Dist. err.} $\downarrow$ & \textbf{Loc. bin} $\uparrow$ & \textbf{Dist. bin} $\uparrow$ & \textbf{Start} $\downarrow$ & \textbf{End} $\downarrow$ & \textbf{tIoU} $\uparrow$ \\
& ($^\circ$) & ($^\circ$) & (m) & (\%) & (\%) & (\%) & (s) & (s) & \\
\midrule
\basemodel{} (vanilla) & 48.0 & 26.4 & 4.16 & 254.0 & 28.8 & 37.5 & 3.96 & 7.19 & 0.13 \\
\basemodel{} (fine-tuned) & 88.2 & 28.3 & 1.36 & 44.0 & 38.2 & 60.1 & 0.26 & 0.34 & 0.90 \\
SELDNet $+$ \basemodel{} & 80.2 & 27.4 & 1.28 & 43.8 & 42.1 & 60.7 & 2.04 & 2.19 & 0.61 \\
\rowcolor{methodrow}\method{} \textit{(ours)} & 17.5 & 14.8 & 1.16 & 39.5 & 82.8 & 64.5 & 0.19 & 0.24 & 0.91 \\
\bottomrule
\end{tabular}}

\vspace{0.35em}
\begin{tabular}{@{}lrrrr@{}}
\toprule
\textbf{Model} & \textbf{Detection acc.} $\uparrow$ & \textbf{Precision} $\uparrow$ & \textbf{Recall} $\uparrow$ & \textbf{F1} $\uparrow$ \\
& (\%) & (\%) & (\%) & (\%) \\
\midrule
\basemodel{} (vanilla) & 22.9 & 22.4 & 7.2 & 10.9 \\
\basemodel{} (fine-tuned) & 63.9 & 66.4 & 94.3 & 78.0 \\
SELDNet $+$ \basemodel{} & 70.8 & 72.6 & 80.3 & 76.2 \\
\rowcolor{methodrow}\method{} \textit{(ours)} & 93.0 & 93.9 & 95.9 & 94.9 \\
\bottomrule
\end{tabular}
\end{table*}

\begin{table*}[t]
\centering
\scriptsize
\caption{Complete Task~3 (Holistic Captioning) results on the evaluation set. Spatial metrics evaluate external sources; temporal and transcription metrics evaluate all applicable matched events. $N$ = total evaluated examples; Format = format adherence; Halluc. = external-source hallucination rate; Missing = missing external source rate; Type acc. = source-type accuracy; Azim. = azimuth MAE; Elev. = elevation MAE; Dist. = distance MAE; Dist.\ err. = percentage distance error; Loc.\ bin = location-bin accuracy; Dist.\ bin = distance-bin accuracy; Start = start-time MAE; End = end-time MAE; tIoU = temporal intersection over union; WER = word error rate; Wearer acc. = example-level wearer-detection accuracy; Precision = precision; Recall = recall; F1 = F1 score.}
\label{tab:evaluation_set_task3_complete}
\resizebox{0.8\linewidth}{!}{
\begin{tabular}{@{}lrrrrr@{}}
\toprule
\textbf{Model} & $\mathbf{N}$ & \textbf{Format} $\uparrow$ & \textbf{Halluc.} $\downarrow$ & \textbf{Missing} $\downarrow$ & \textbf{Type acc.} $\uparrow$ \\
& & (\%) & (\%) & (\%) & (\%) \\
\midrule
\basemodel{} (vanilla) & 33,977 & 100.0 & 6.7 & 75.4 & 60.8 \\
\basemodel{} (fine-tuned) & 33,977 & 100.0 & 18.5 & 7.9 & 62.4 \\
SELDNet $+$ \basemodel{} & 33,977 & 100.0 & 50.1 & 20.0 & 56.3 \\
\rowcolor{methodrow}\method{} \textit{(ours)} & 33,977 & 100.0 & 5.8 & 4.9 & 63.8 \\
\bottomrule
\end{tabular}}

\vspace{0.35em}
\resizebox{\linewidth}{!}{
\begin{tabular}{@{}lrrrrrrrrrr@{}}
\toprule
\textbf{Model} & \textbf{Azim.} $\downarrow$ & \textbf{Elev.} $\downarrow$ & \textbf{Dist.} $\downarrow$ & \textbf{Dist. err.} $\downarrow$ & \textbf{Loc. bin} $\uparrow$ & \textbf{Dist. bin} $\uparrow$ & \textbf{Start} $\downarrow$ & \textbf{End} $\downarrow$ & \textbf{tIoU} $\uparrow$ & \textbf{WER} $\downarrow$ \\
& ($^\circ$) & ($^\circ$) & (m) & (\%) & (\%) & (\%) & (s) & (s) & & (\%) \\
\midrule
\basemodel{} (vanilla) & 65.3 & 27.6 & 1.71 & 61.1 & 21.9 & 29.7 & 3.11 & 4.16 & 0.37 & 24.5 \\
\basemodel{} (fine-tuned) & 86.7 & 27.6 & 1.25 & 46.6 & 32.6 & 57.3 & 0.16 & 0.40 & 0.92 & 21.6 \\
SELDNet $+$ \basemodel{} & 80.3 & 27.4 & 1.22 & 42.4 & 42.0 & 61.2 & 1.99 & 2.33 & 0.58 & 19.3 \\
\rowcolor{methodrow}\method{} \textit{(ours)} & 14.2 & 12.8 & 1.06 & 36.8 & 83.9 & 66.8 & 0.10 & 0.17 & 0.94 & 17.8 \\
\bottomrule
\end{tabular}}

\vspace{0.35em}
\begin{tabular}{@{}lrrrr@{}}
\toprule
\textbf{Model} & \textbf{Wearer acc.} $\uparrow$ & \textbf{Precision} $\uparrow$ & \textbf{Recall} $\uparrow$ & \textbf{F1} $\uparrow$ \\
& (\%) & (\%) & (\%) & (\%) \\
\midrule
\basemodel{} (vanilla) & 62.2 & 66.0 & 58.9 & 62.3 \\
\basemodel{} (fine-tuned) & 96.3 & 99.0 & 94.1 & 96.4 \\
SELDNet $+$ \basemodel{} & 96.1 & 93.4 & 99.6 & 96.4 \\
\rowcolor{methodrow}\method{} \textit{(ours)} & 99.9 & 100.0 & 99.9 & 99.9 \\
\bottomrule
\end{tabular}
\end{table*}

\begin{table*}[t]
\centering
\scriptsize
\caption{Complete Task~4 (Role-Aware Transcription) results on the evaluation set. Spatial metrics evaluate matched bystander speech; temporal and transcription metrics evaluate all applicable matched speech. $N$ = total evaluated examples; Format = format adherence; Halluc. = bystander-speech hallucination rate; Missing = missing external source rate; Azim. = azimuth MAE; Elev. = elevation MAE; Dist. = distance MAE; Dist.\ err. = percentage distance error; Loc.\ bin = location-bin accuracy; Dist.\ bin = distance-bin accuracy; Start = start-time MAE; End = end-time MAE; tIoU = temporal intersection over union; WER = word error rate; Wearer acc. = example-level wearer-detection accuracy; Precision = precision; Recall = recall; F1 = F1 score.}
\label{tab:evaluation_set_task4_complete}
\resizebox{0.8\linewidth}{!}{
\begin{tabular}{@{}lrrrr@{}}
\toprule
\textbf{Model} & $\mathbf{N}$ & \textbf{Format} $\uparrow$ & \textbf{Halluc.} $\downarrow$ & \textbf{Missing} $\downarrow$ \\
& & (\%) & (\%) & (\%) \\
\midrule
\basemodel{} (vanilla) & 25,876 & 100.0 & 15.8 & 45.0 \\
\basemodel{} (fine-tuned) & 25,876 & 100.0 & 15.6 & 3.4 \\
SELDNet $+$ \basemodel{} & 25,876 & 100.0 & 10.9 & 16.1 \\
\rowcolor{methodrow}\method{} \textit{(ours)} & 25,876 & 100.0 & 4.5 & 1.8 \\
\bottomrule
\end{tabular}}

\vspace{0.35em}
\resizebox{\linewidth}{!}{
\begin{tabular}{@{}lrrrrrrrrrr@{}}
\toprule
\textbf{Model} & \textbf{Azim.} $\downarrow$ & \textbf{Elev.} $\downarrow$ & \textbf{Dist.} $\downarrow$ & \textbf{Dist. err.} $\downarrow$ & \textbf{Loc. bin} $\uparrow$ & \textbf{Dist. bin} $\uparrow$ & \textbf{Start} $\downarrow$ & \textbf{End} $\downarrow$ & \textbf{tIoU} $\uparrow$ & \textbf{WER} $\downarrow$ \\
& ($^\circ$) & ($^\circ$) & (m) & (\%) & (\%) & (\%) & (s) & (s) & & (\%) \\
\midrule
\basemodel{} (vanilla) & 73.4 & 27.2 & 1.65 & 63.9 & 18.9 & 38.4 & 3.03 & 3.91 & 0.39 & 27.7 \\
\basemodel{} (fine-tuned) & 87.2 & 27.3 & 1.19 & 47.8 & 35.9 & 54.5 & 0.14 & 0.41 & 0.93 & 21.8 \\
SELDNet $+$ \basemodel{} & 80.1 & 27.0 & 1.18 & 37.9 & 42.2 & 64.8 & 2.17 & 2.42 & 0.55 & 17.7 \\
\rowcolor{methodrow}\method{} \textit{(ours)} & 13.7 & 11.9 & 1.00 & 35.2 & 85.2 & 68.9 & 0.08 & 0.15 & 0.95 & 17.8 \\
\bottomrule
\end{tabular}}

\vspace{0.35em}
\begin{tabular}{@{}lrrrr@{}}
\toprule
\textbf{Model} & \textbf{Wearer acc.} $\uparrow$ & \textbf{Precision} $\uparrow$ & \textbf{Recall} $\uparrow$ & \textbf{F1} $\uparrow$ \\
& (\%) & (\%) & (\%) & (\%) \\
\midrule
\basemodel{} (vanilla) & 51.4 & 67.4 & 55.6 & 61.0 \\
\basemodel{} (fine-tuned) & 95.0 & 99.0 & 93.6 & 96.2 \\
SELDNet $+$ \basemodel{} & 95.3 & 93.9 & 99.6 & 96.7 \\
\rowcolor{methodrow}\method{} \textit{(ours)} & 99.9 & 100.0 & 100.0 & 100.0 \\
\bottomrule
\end{tabular}
\end{table*}

\begin{table*}[t]
\centering
\scriptsize
\caption{Complete Task~5 (Targeted Wearer Speech Transcription) results on the evaluation set. $N$ = total evaluated examples; Format = format adherence; Start = start-time MAE; End = end-time MAE; tIoU = temporal intersection over union; WER = word error rate; Wearer acc. = example-level wearer-detection accuracy; Precision = precision; Recall = recall; F1 = F1 score.}
\label{tab:evaluation_set_task5_complete}
\resizebox{\linewidth}{!}{
\begin{tabular}{@{}lrrrrrrrrrr@{}}
\toprule
\textbf{Model} & $\mathbf{N}$ & \textbf{Format} $\uparrow$ & \textbf{Start} $\downarrow$ & \textbf{End} $\downarrow$ & \textbf{tIoU} $\uparrow$ & \textbf{WER} $\downarrow$ & \textbf{Wearer acc.} $\uparrow$ & \textbf{Precision} $\uparrow$ & \textbf{Recall} $\uparrow$ & \textbf{F1} $\uparrow$ \\
& & (\%) & (s) & (s) & & (\%) & (\%) & (\%) & (\%) & (\%) \\
\midrule
\basemodel{} (vanilla) & 17,372 & 91.8 & 3.20 & 4.44 & 0.37 & 38.5 & 77.8 & 100.0 & 77.8 & 87.5 \\
\basemodel{} (fine-tuned) & 17,372 & 100.0 & 0.05 & 0.34 & 0.95 & 8.9 & 100.0 & 100.0 & 100.0 & 100.0 \\
SELDNet $+$ \basemodel{} & 17,372 & 100.0 & 2.27 & 2.34 & 0.53 & 6.5 & 100.0 & 100.0 & 100.0 & 100.0 \\
\rowcolor{methodrow}\method{} \textit{(ours)} & 17,372 & 100.0 & 0.02 & 0.04 & 0.98 & 5.8 & 100.0 & 100.0 & 100.0 & 100.0 \\
\bottomrule
\end{tabular}}
\end{table*}

\begin{table*}[t]
\centering
\scriptsize
\caption{Complete Task~6 (Targeted External Speech Transcription) results on the evaluation set. $N$ = total evaluated examples; Format = format adherence; Halluc. = external-speech hallucination rate; Missing = missing external source rate; Azim. = azimuth MAE; Elev. = elevation MAE; Dist. = distance MAE; Dist.\ err. = percentage distance error; Loc.\ bin = location-bin accuracy; Dist.\ bin = distance-bin accuracy; Start = start-time MAE; End = end-time MAE; tIoU = temporal intersection over union; WER = word error rate; Detection acc. = source-level query-detection accuracy; Precision = precision; Recall = recall; F1 = F1 score.}
\label{tab:evaluation_set_task6_complete}
\resizebox{0.8\linewidth}{!}{
\begin{tabular}{@{}lrrrr@{}}
\toprule
\textbf{Model} & $\mathbf{N}$ & \textbf{Format} $\uparrow$ & \textbf{Halluc.} $\downarrow$ & \textbf{Missing} $\downarrow$ \\
& & (\%) & (\%) & (\%) \\
\midrule
\basemodel{} (vanilla) & 14,921 & 100.0 & 25.8 & 64.2 \\
\basemodel{} (fine-tuned) & 14,921 & 100.0 & 27.2 & 0.5 \\
SELDNet $+$ \basemodel{} & 14,921 & 100.0 & 13.4 & 6.3 \\
\rowcolor{methodrow}\method{} \textit{(ours)} & 14,921 & 100.0 & 8.0 & 1.2 \\
\bottomrule
\end{tabular}}

\vspace{0.35em}
\resizebox{\linewidth}{!}{
\begin{tabular}{@{}lrrrrrrrrrr@{}}
\toprule
\textbf{Model} & \textbf{Azim.} $\downarrow$ & \textbf{Elev.} $\downarrow$ & \textbf{Dist.} $\downarrow$ & \textbf{Dist. err.} $\downarrow$ & \textbf{Loc. bin} $\uparrow$ & \textbf{Dist. bin} $\uparrow$ & \textbf{Start} $\downarrow$ & \textbf{End} $\downarrow$ & \textbf{tIoU} $\uparrow$ & \textbf{WER} $\downarrow$ \\
& ($^\circ$) & ($^\circ$) & (m) & (\%) & (\%) & (\%) & (s) & (s) & & (\%) \\
\midrule
\basemodel{} (vanilla) & 40.7 & 25.9 & 1.85 & 89.1 & 100.0 & 39.6 & 3.31 & 4.15 & 0.39 & 49.0 \\
\basemodel{} (fine-tuned) & 31.7 & 26.5 & 1.25 & 50.0 & 100.0 & 53.7 & 0.28 & 0.71 & 0.88 & 38.1 \\
SELDNet $+$ \basemodel{} & 34.9 & 26.3 & 1.20 & 39.2 & 100.0 & 63.7 & 2.39 & 2.90 & 0.50 & 36.2 \\
\rowcolor{methodrow}\method{} \textit{(ours)} & 10.5 & 11.8 & 1.03 & 35.6 & 100.0 & 69.3 & 0.18 & 0.32 & 0.91 & 32.5 \\
\bottomrule
\end{tabular}}

\vspace{0.35em}
\begin{tabular}{@{}lrrrr@{}}
\toprule
\textbf{Model} & \textbf{Detection acc.} $\uparrow$ & \textbf{Precision} $\uparrow$ & \textbf{Recall} $\uparrow$ & \textbf{F1} $\uparrow$ \\
& (\%) & (\%) & (\%) & (\%) \\
\midrule
\basemodel{} (vanilla) & 24.6 & 31.0 & 8.4 & 13.2 \\
\basemodel{} (fine-tuned) & 74.8 & 71.8 & 99.3 & 83.4 \\
SELDNet $+$ \basemodel{} & 81.3 & 83.1 & 91.1 & 86.9 \\
\rowcolor{methodrow}\method{} \textit{(ours)} & 91.6 & 90.3 & 98.3 & 94.1 \\
\bottomrule
\end{tabular}
\end{table*}

\section{AI Usage}
\label{appendix:ai_usage}
We use LLMs to help with the writing of the paper in terms of: (1) grammar check, and (2) occasionally choosing the best word in writing, (3) rewrite few sentences for better clarity and space management. We also use LLMs to for literature discovery. We use LLMs as part of data curation in our research as discussed in our method section, in a similar way as many other LLM-related research papers.

\end{document}